\documentstyle[psfig,subfigure,astrobib,epsfig]{mn}
\title[H110$\alpha$ recombination-line emission and 4.8-GHz continuum
emission in the Carina Nebula]
{H110$\alpha$ recombination-line emission and 4.8-GHz continuum
emission in the Carina Nebula}
\author[Brooks, Storey, Whiteoak]
{K. J.~Brooks$^{1}$\thanks{Current address: European Southern Observatory,
Casilla 19001, Santiago 19, Chile (kbrooks@eso.org).}, J. W. V. Storey$^{1}$ and J. B. Whiteoak$^{2}$\\
$^1$ Department of Physics, University of New South
Wales, Sydney 2052, NSW Australia\\
$^2$Australia Telescope National Facility, CSIRO, PO Box 76, Epping 2121,
NSW Australia\\
}
\newcommand{\kms}{km\,s$^{-1}$}
\newcommand{\Msun}{M$_{\odot}$}
\newcommand{\Lsun}{L$_{\odot}$}
\newcommand{\HII}{$\mathrm{H\,{\scriptstyle II}}\,$}

\begin{document}
%\label{firstpage}
\maketitle
 
\begin{abstract}

We present results from observations of H110$\alpha$ recombination-line
emission at 4.874 GHz and the related 4.8-GHz continuum emission towards
the Carina Nebula using the Australia Telescope Compact Array. These data
provide information on the velocity, morphology and excitation parameters
of the ionized gas associated with the two bright \HII\ regions within the
nebula, Car I and Car II. They are consistent with both Car I and Car II
being expanding ionization fronts arising from the massive star clusters
Trumpler 14 and Trumpler 16, respectively. The overall continuum emission
distribution at 4.8 GHz is similar to that at lower frequencies. For Car I,
two compact sources are revealed that are likely to be young \HII\ regions
associated with triggered star formation. These results provide the first
evidence of ongoing star formation in the northern region of the nebula. A
close association between Car~I and the molecular gas is consistent with a
scenario in which Car~I is currently carving out a cavity within the
northern molecular cloud. The complicated kinematics associated with Car~II
point to expansion from at least two different centres. All that is left of
the molecular cloud in this region are clumps of dense gas and dust which
are likely to be responsible for shaping the striking morphology of the
Car~II components.
\end{abstract}


\section{Introduction}

It is well known that the radiation fields and stellar winds of massive
stars can drastically affect the physical conditions, structure, and
chemistry of the giant molecular cloud (GMC) from which they formed. It is
also thought that massive stars are at least partly responsible for
triggering further star formation within a GMC. The details of this
interaction, however, are not well understood and additional detailed study
of massive star-forming regions is needed.

The Carina Nebula is a spectacular star-forming region containing some of
the most massive stars known in our Galaxy. The two most influential
clusters of the nebula, Trumpler~14 and 16 (hereafter Tr~14 and Tr~16),
contain 6 O3 type stars (Feinstein 1995).  Tr~16 also contains the
well-known star $\eta$~Car. The nebula is part of a GMC complex that
extends over 130 pc with a mass in excess of 5$\times$10$^{5}$ \Msun\
\cite{Grabelsky88}. It consists of two main emission regions identified as
the northern and southern clouds (\citeNP{deGraauw81}, \citeNP{Whiteoak84},
\citeNP{Cox952} and \citeNP{Brooks98}).  The region between these two
clouds is centred on the Keyhole Nebula and here the molecular gas exists
in several dense clumps of typical mass 10 \Msun\ (\citeNP{Cox951} and
\citeNP{Brooks00}).

Thermal radio continuum emission has been detected over a large part of the
nebula, with two bright concentrations named Car~I and Car~II
\cite{Gardner68}. These two \HII\ regions are ionized by Tr~14 and Tr~16
respectively, and contain bright ionization fronts (\citeNP{Retallack83}
and \citeNP{Whiteoak94}). Car~I is located in the northern part of the
nebula and overlaps with both the optical nebulosity and the western dust
lane. Car~II is located towards the Keyhole Nebula and traces the bright
optical emission filaments present in this region. The large-scale dynamics
measured across the nebula are extremely complex, and some interpretations
involve merging spiral arms \cite{Tateyama91}, rotating neutral clouds
\cite{Meaburn84} or old \HII\ regions \cite{Cersosimo84}. On a smaller
scale, the dynamics of the ionized gas in the vicinity of Car~II are
indicative of an expanding shell that is likely the result of the strong
stellar winds from the stellar members of Tr~16, in particular $\eta$~Car
(\citeNP{Deharveng75}, \citeNP{Huchtmeier75} and \citeNP{Cox952}).

The picture emerging for the Carina Nebula is one in which the extreme
radiation flux and stellar winds from Tr~14 and Tr~16 have carved out large
cavities within the GMC \cite{Brooks00t}. The close proximity of Tr~14 to
the northern molecular cloud  suggests that such a process is
still taking place here.  Evidence for
ongoing star-formation in the Carina Nebula is scarce. Only a small number
of sites have been identified in the southern molecular cloud
(\citeNP{Megeath96} and \citeNP{Smith00}) and little is known about the
processes taking place in the northern cloud.

In this paper we investigate further the ionized gas associated with Car~I
and Car~II. We have made observations of H110$\alpha$ (4.874~GHz)
recombination-line emission and the corresponding 4.8-GHz continuum
emission using the Australia Telescope Compact Array (ATCA)\footnote{The
Australia Telescope Compact Array is funded by the Commonwealth of
Australia for operation as a National Facility managed by CSIRO}. These new
data not only support the presence of expanding gas shells, but also
provide details on the small-scale distribution of the ionization fronts
and reveal continuum emission sources resembling compact \HII\ regions.

Hereafter we will adopt the popular view (e.g. \citeNP{Tovmassian95},
\citeNP{Walborn95} and \citeNP{Davidson97}) that Tr~14 and Tr~16 are at a
common distance of 2.2 $\pm$ 0.2 kpc and that Tr~14 is younger ($1\times
10^6$~yr) than Tr~16 ($3\times 10^6$~yr).

\section{Observations}

The ATCA observations were carried out between 1996 April and 1996 October.
Details of the instrument are given by \citeN{Frater92}. Three different
array configurations were utilised (375, 750D and 1.5A), each for a
duration of about 12 hours.  The correlator was configured so that both
line and continuum observations were obtained simultaneously. For the line
measurements, a bandwidth of 8~MHz was used with 512 channels, providing a
velocity resolution of 0.96~\kms. A bandwidth of 128~MHz with 32 channels
was used for the continuum observations. At 4.8~GHz the half-power primary
beam of the ATCA is 10~arcmin. Primary-beam attenuation corrections were
made to all of the final images.  Two pointing centres were used:
RA(B1950)~$ = 10^{h}41^{m}36^{s}$, Dec(B1950)~$ =
-59^{\circ}19\arcmin00\arcsec$ for Car~I and RA(B1950)~$ =
10^{h}42^{m}54^{s}$, Dec(B1950)~$ = -59^{\circ}23\arcmin00\arcsec$ for
Car~II. A cycle of twenty-minute observations of each pointing centre
bracketed by 3-minute observations of a phase calibrator, either
PKS~1215-457 or PKS~1039-47, was used. These calibrators were also used to
calibrate the spectral bandpass. The flux  was calibrated by observing
PKS~1934-638 (the ATCA primary calibrator), for which values of 5.73~Jy at
4.874~GHz and 5.83~Jy at 4.8~GHz were adopted. Data were edited and
calibrated according to standard techniques using the software package
{\sc miriad} (see \citeNP{Sault95} and references therein).

The continuum emission associated with $\eta$~Car varied both in amplitude
and spatial extent throughout the 1996 observations (see
\citeNP{Duncan97}). In our data we detected a flux density variation between
0.1 and 0.2~Jy ~beam$^{-1}$ which affected the de-convolution of the Car~II
data. The image quality of Car~II was optimised by removing a suitable
point source model for $\eta$~Car from each set of \mbox{{\sc u-v}}
data. However, low-amplitude residuals remain towards the position of
$\eta$~Car.

The continuum images were formed using the techniques of multi-frequency
synthesis. Both uniform and natural weighting were used to produce two
images, the latter of which was used for obtaining line-to-continuum
measurements.  For the uniform-weighted images, the Car~I and Car~II data
were imaged separately and deconvolved using the maximum entropy algorithm,
{\sc mem}. Diffraction-limited restoring beams of $7.8 \times 6.5$
arcsec$^2$ and $8.6 \times 6.6$~arcsec$^2$ were used for the images of
Car~I and Car~II respectively. The final images have an rms noise of
1~mJy~beam$^{-1}$. For the natural-weighted images, the sensitivity was
optimised by combining the Car~I and Car~II data as a mosaic. This single
mosaic image was deconvolved using the {\sc clean} algorithm and then
restored with the diffraction-limited beam of $28.3 \times
26.0$~arcsec$^2$. The final image is more sensitive to the missing zero
spacings and has a flux density uncertainty of 50~mJy~beam$^{-1}$.

Continuum emission was subtracted from the line data in the {\sc u-v}
plane. A data cube (RA, Dec, velocity) was formed by combining the Car~I
and Car~II data as a mosaic using natural weighting. Each channel image was
deconvolved using the {\sc clean} algorithm and restored
with a diffraction-limited beam of $28.3 \times 26.0$~arcsec$^2$. The rms
noise value of an emission-free channel image is 5.4~mJy~beam$^{-1}$.

\section{Results}
\subsection{4.8-GHz continuum emission}

\begin{figure*}
\psfig{file=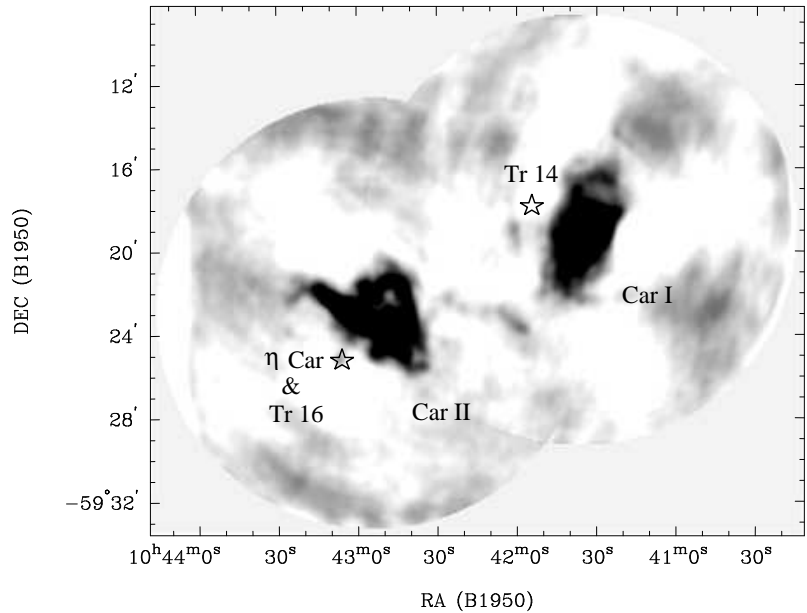,width=0.6\textwidth}
\caption{\label{h110_cont_l}Natural-weighted 4.8-GHz continuum image of Car
I and Car~II with grey-scale intensity levels ranging from $-0.01$ to
0.2~Jy~beam$^{-1}$. (See Fig.~\ref{h110_regions} for a contour
representation.) The stars mark the approximate location of Tr~16 (taken
from the position of $\eta$~Car) and Tr~14.}
\end{figure*}

The natural-weighted continuum image (Fig.~\ref{h110_cont_l}) illustrates
very well the spatial relation between Car~I and Car~II. Their overall
emission features agree with previous continuum observations made by
\citeN{Whiteoak94} using the Molongolo Observatory Synthesis Telescope
(MOST) at 0.843~GHz with a resolution of 30~arcsec.  We have adopted the
same nomenclature.  Prominent in the
grey-scale representation is the non-uniformity of the background noise
caused by missing zero spacings. The string-like emission feature extending
between Car~I and Car~II, at the centre of the image, corresponds to a
similar feature in the MOST image. The same is true for the faint clump to
the east of Car~I.

\begin{figure*}
\psfig{bbllx=24pt,bblly=460pt,bburx=560pt,bbury=710pt,file=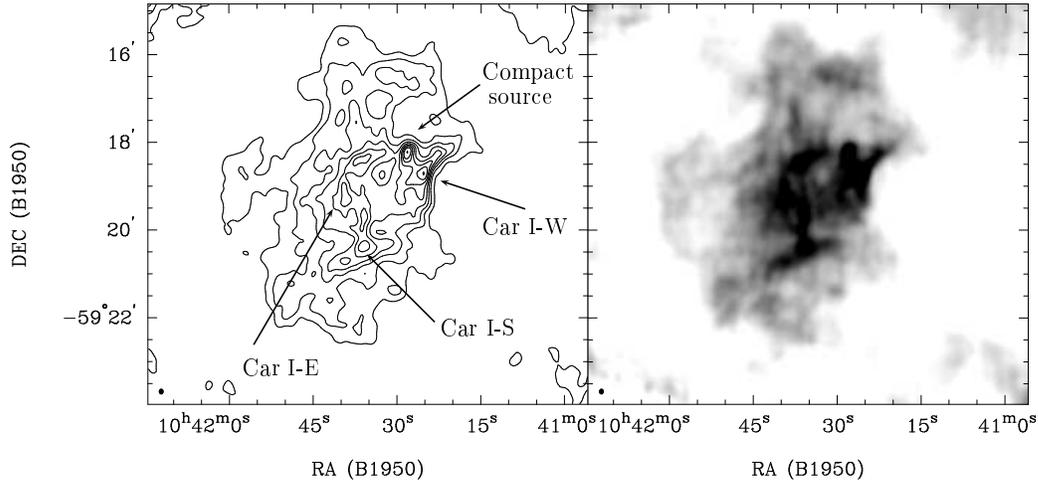,width=0.8\textwidth,silent=}
\caption{\label{h110_cont_cari}Uniform-weighted 4.8-GHz continuum image of
Car~I. {\it Left} --- Contour representation with levels 0.006, 0.012,
0.018, 0.024, 0.030, 0.036 0.042, 0.048, and 0.056~Jy~beam$^{-1}$. {\it
Right} --- Grey-scale representation for intensity levels ranging from
0.002 to 0.032~Jy beam$^{-1}$. The beamsize is illustrated in the lower
left of each image.}
\end{figure*}

The uniform-weighted continuum image of Car~I is shown in
Fig.~\ref{h110_cont_cari}. The overall emission distribution is elongated
almost north--south over $5 \times 8$~arcmin$^2$. Superimposed on this
extended emission are a number of striking features which were first noted
(but without detail) by \citeN{Whiteoak94}. These features are enhanced in
the grey-scale representation.  Car~I-E is the more eastern feature and is
located near the centre of the extended emission. It forms an arc of length
about 2~arcmin that bends towards the west and contains bright filaments
and knots of emission. Car~I-S is located at the southern end of
Car~I-E. Here the emission extends over 1.5~arcmin in the
northwest--southeast direction with a central emission knot that resembles
a compact source. Car~I-W was first described by \citeN{Whiteoak94} as an
arc that bends eastwards, forming a partial ring feature with
\mbox{Car~I-E}. However, the higher resolution images presented here
indicate that Car~I-W is made up of two main components --- a compact
source located in the northeast and an arc-shaped feature in the west
(hereafter redefined as Car~I-W). Car~I-W is well defined over 2~arcmin and
bends towards the west, in the same direction as Car~I-E. Its western edge
is very sharp and marks the western boundary of the overall extended
emission. 

%\begin{sidewaystable} 
\begin{table*}
\begin{minipage}{\textwidth}
\caption{\label{h110_table1}Total integrated fluxes ($S_{Total}$) and
physical parameters derived from the uniform-weighted 4.8-GHz continuum
images. The listed positions correspond to the location of the peak flux
density ($S_{Peak}$) of each feature. Their sizes have been defined using a
contour level of 0.03~Jy for Car~I and 0.02~Jy for Car~II. Physical
parameters include, optical depth $\tau$, emission measure $EM$ and
the total number of ionizing photons per second $N_{L}$.}
\begin{tabular}{@{}lcccccccc}
\hline
Source  & RA (B1950)
        & Dec (B1950)         
        & $S_{Peak}$\footnote{With an uncertainty of $\pm 0.003$~Jy~beam$^{-1}$}

        & $S_{Total}$   
        & Size 
	& $\tau$
	& $EM$
	& $N_{L}$
\\
        &  ~$^h$ ~~$^m$ ~~$^s$ 
        & \degr  ~~$'$ ~~\arcsec
        & Jy~beam$^{-1}$ 
        & Jy   
	& $\arcmin \times \arcmin$
	& $\times 10^{-2}$
 	& $ \times10^5$ pc cm$^{-6}$
        & $\times10^{48}$ s$^{-1}$
\\
\hline
\it{Car~I}      &       &              &       &       & 	& & &\\
Total   
        & 10 41 28.2   
        & $-59$ 18 14  
        & 0.056 
        & $23.4\pm0.1$  
        & $5 \times 8 $
	&
	&
	& $12 \pm 1$ 
\\
Car~I-E       
        & 10 41 35.7  
        & $-59$ 19 42   
        & 0.039
        & $2.19\pm0.03$  
        &
	& $0.6 \pm 0.2$
	& $2.8 \pm 0.4$
	& $1.2 \pm 0.1$
\\

Car~I-S 
        & 10 41 36.0     
        & $-59$ 20 24  
        & 0.040
        & $0.28\pm0.01$
        & $0.5 \times 0.5$
	& $0.7 \pm 0.2$
	& $2.9 \pm 0.4$
	& $0.15 \pm 0.02$ 
\\
Car~I-W     
        & 10 41 25.3
        & $-59$ 18 42   
        & 0.048
        & $1.07\pm0.03$
        &
	& $0.8 \pm 0.2$
        & $3.4 \pm 0.4$ 
        & $0.59 \pm 0.05$
\\   

Compact     
        & 10 41 28.1   
        & $-59$ 18 14  
        & 0.056 
        & $0.61\pm0.01$    
        & $0.3 \times 0.4$
	& $1.0 \pm 0.2$
        & $4.0 \pm 0.4$
        & $0.33 \pm 0.03$
\\
\it{Car~II}     &             &       &       &	&&&\\
Total 
        & 10 43 07.1 
        & $-59$ 22 34   
        & 0.076
        & $22.93\pm0.09$  
        & $6 \times 4.5$
	&
	&
	&$1.23 \pm 0.08$
\\
Car~II-N   
        & 10 42 42.0   
        & $-59$ 21 44   
        & 0.061
        & $7.10\pm0.05$   
        &
	& $1.0 \pm 0.3$
        & $3.9 \pm 0.4$ 
        & $3.8 \pm 0.3$
\\
Car~II-W        
        & 10 42 42.0   
        & $-59$ 23 54  
        & 0.074
        & $5.08 \pm 0.04$   
        & $2 \times 1$
	& $1.2 \pm 0.3$
        & $4.7 \pm 0.5$
        & $2.7 \pm 0.3$
\\
Car~II-E        
        & 10 43 07.1    
        & $-59$ 22 34   
        & 0.076
        & $7.01 \pm 0.05$   
        & $3\times 1$
	& $1.2 \pm 0.2$
        & $4.8 \pm 0.5$
        & $3.8 \pm 0.4$
\\
\hline 
\end{tabular}
\end{minipage}
\end{table*}
%\end{sidewaystable}
 
\begin{figure*}
\psfig{bbllx=14pt,bblly=460pt,bburx=575pt,bbury=733pt,file=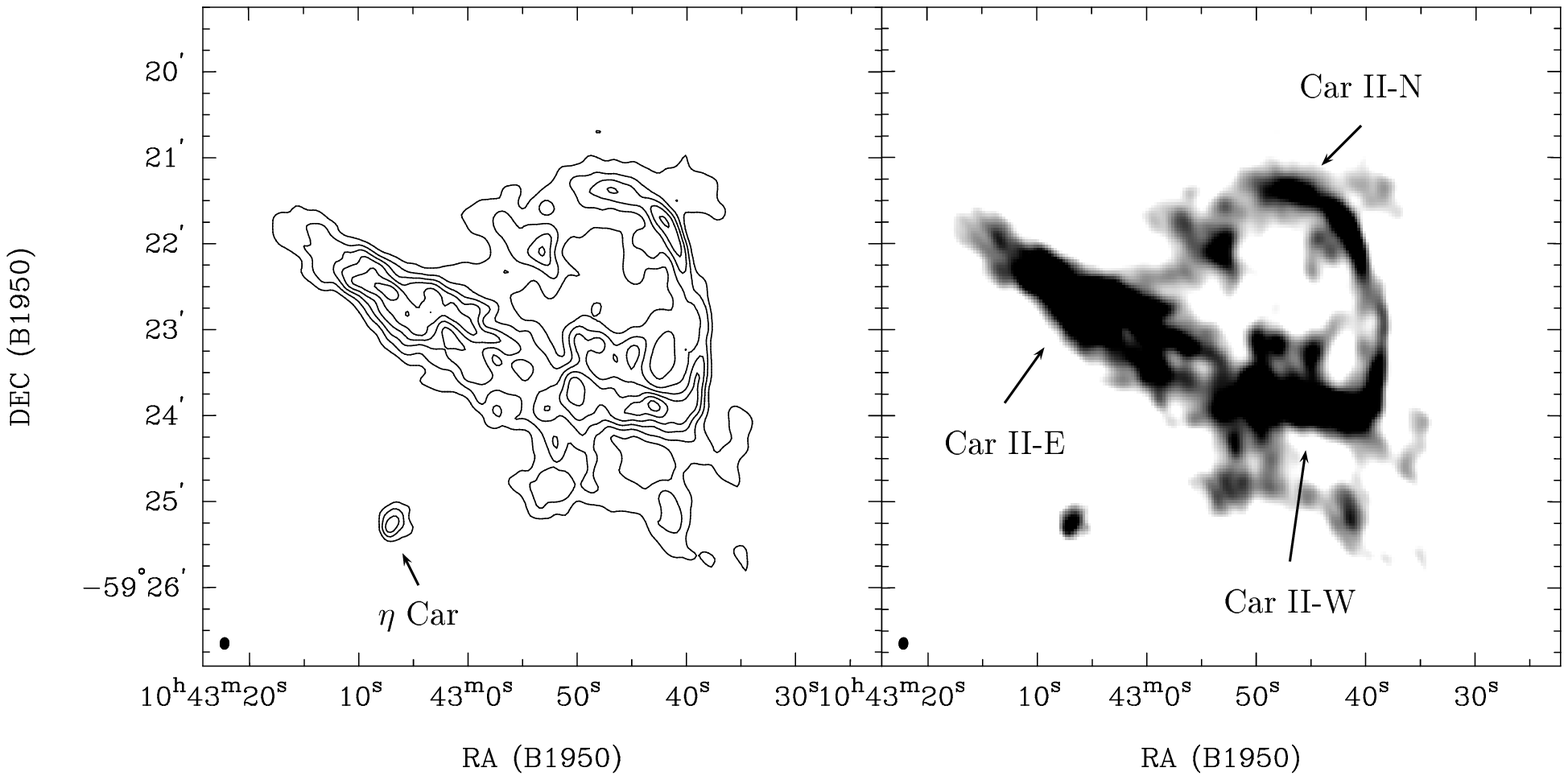,width=0.8\textwidth,silent=}
\caption{\label{h110_cont_carii}Uniform-weighted 4.8~GHz continuum
emission image of Car~II. {\it Left} --- Contour representation with levels
0.02, 0.03, 0.04, 0.05, 0.06, 0.07~Jy~beam$^{-1}$. {\it Right} ---
Grey-scale representation, ranging from 0.02 to 0.04~Jy~beam$^{-1}$. The
beamsize is illustrated in the lower left of
each image. Note that the intensities for $\eta$~Car are only the residual
caused by subtraction of the source from the data.}
\end{figure*}

Fig.~\ref{h110_cont_carii} shows the uniform-weighted continuum
image of Car~II. There are three well-defined, bright emission components,
all of which were first identified in a study by \citeN{Retallack83} using
the Fleurs Synthesis Telescope at 1.415~GHz with a resolution of
1~arcmin. The higher resolution data obtained here
reveal that they are composed of a number of bright emission filaments
and knots.  They constitute a prominent ring-like structure of diameter
2~arcmin.  One component, Car~II-N, is in the shape of an arc and
forms more than half of the ring. The two other components, Car~II-E and
Car~II-W, form a linear or bar-like feature that extends over
5~arcmin in the northeast--southwest direction. 

Parameters measured from the main continuum components of Car~I and Car~II
are listed in Table~\ref{h110_table1}. All the calculations have been based
on the assumption that Car~I and Car~II are ionization bounded with uniform
density and have optically thin continuum emission. A value of 6000 ($\pm$
1000~K) was used for the electron temperature. This represents values for
the strong continuum features derived from the H110$\alpha$ emission (see
later discussion).  The parameters include 4.8-GHz continuum optical depth,
emission measure (calculated according to \citeNP{Mezger67}) and ionizing
photon flux per second (calculated according to
\citeNP{Molinari98}). Estimates for the electron density and mass were
obtained for the two compact sources in Car~I by assuming that they are
near-spherical (see Table~\ref{contparam}). Using the results from
\citeN{Panagia73}, the spectral types of a single ionizing star for each of
these two sources were also estimated.

%\begin{sidewaystable}
\begin{table}
\begin{minipage}{\columnwidth}
\caption{\label{contparam} Additional physical parameters derived from the
compact emission features in Car~I. Parameters include source size
$\theta_s$, electron density $N_{e}$, mass $M$ and the single exciting
star type. }
\begin{tabular}{@{}lccccc}
\hline
Region  & $\theta_s$
        & $N_e$ 
        & $M$
        & Star
\\
        & arcsec
        & $\times10^{3}$ cm$^{-3}$
        & \Msun
        & type
\\
\hline

Car~I-S         & 44
                & $0.75 \pm 0.05$
                & 4
                & BO
\\

Compact 	& 30
                & $1.08 \pm 0.06$
                & 2 
                & O9.5

\\\hline 
\end{tabular}
\end{minipage} 
\end{table}

\subsection{H110$\alpha$ recombination-line emission}

\begin{figure*}
\psfig{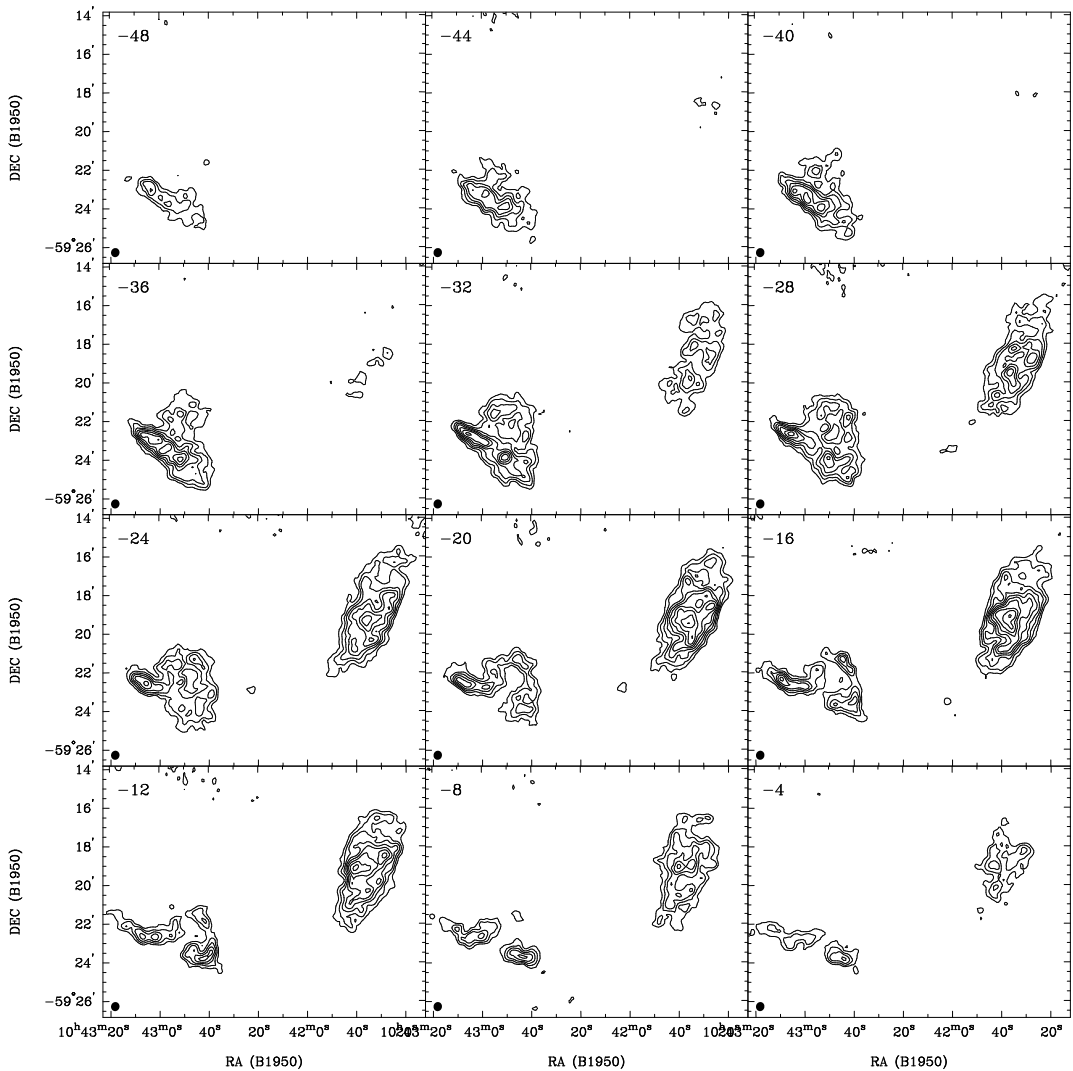}
\caption{\label{h110_chanmaps}A sequence of H110$\alpha$ line images of Car
I and Car~II for velocities ranging from $-48$ to $-4$~\kms, with a
velocity resolution of 4~\kms. The contours are 0.01, 0.015, 0.020, 0.025,
0.030, 0.035, 0.040, 0.045, 0.050~Jy~beam$^{-1}$. For each image the
beamsize is shown lower left and the velocity (in \kms) is shown upper left.}
\end{figure*}

Fig.~\ref{h110_chanmaps} shows channel images of the H110$\alpha$ emission
associated with Car~I and Car~II, covering a velocity range from $-48$ to $-4$ 
\kms. 

The emission associated with Car~I occurs over a velocity range from $-$36
to $-$4~\kms\ and follows the general shape of the continuum emission with
no apparent velocity trends.

The velocity structure of the emission associated with Car~II is more
complicated and covers a wider range from $-$48 to $-$4 \kms. Two main
types of distribution features are evident: one is most prominent near
$-$40~\kms\ and consists of a broad ridge of emission extending
northeast--southwest, corresponding to the continuum emission bar
(Car~II-E and Car~II-W); the other is more prominent near $-$20 \kms\ and
consists of a semi-circular feature, corresponding to Car II-N. A central
hole is evident between velocities of -32 and -24 \kms, suggesting that the
semi-circular feature could be part of a complete ring or shell. At these
velocities there is also emission resembling the distribution of Car~II-E
and Car~II-W. However, it is narrower and clearly separated toward the
centre.

\begin{figure*}
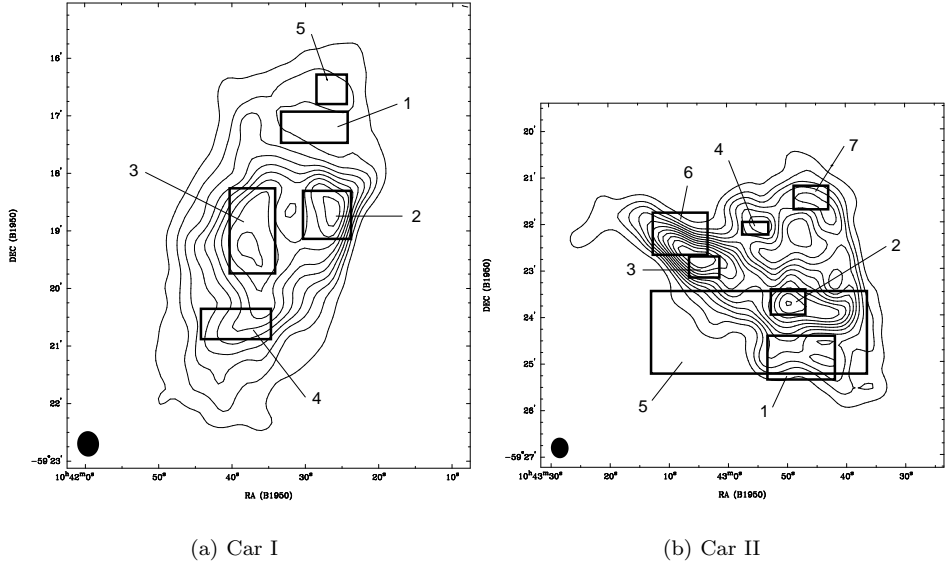

\centering \mbox{\subfigure[Car~I
]{\psfig{file=fig5a.eps,clip=,width=0.35\textwidth}}}
\mbox{\subfigure[Car~II]{\psfig{file=fig5b.eps,clip=,width=0.35\textwidth}}}
\caption{\label{h110_regions}Selected regions in Car~I and Car~II where
integrated line profiles have been obtained. The individual profiles are
shown in Fig.~\ref{h110_spectra1} and Fig.~\ref{h110_spectra2} and their
parameters listed in Table~\ref{h110_params}. The contour images are
analogous to Fig. 1 with levels 0.08, 0.15, 0.2, 0.25, 0.3, 0.35, 0.4,
0.45, 0.5, 0.55, 0.6, 0.65~Jy beam$^{-1}$.}
\end{figure*}

A representative set of line profiles with good signal-to-noise has been
obtained by integrating the emission over each of the Car~I and Car~II
regions as well as several selected regions within them (indicated in
Fig.~\ref{h110_regions}.) Results from Gaussian fits to each of the
profiles are listed in Table~\ref {h110_params}. For the H110$\alpha$
emission, the linewidth contains a component from thermal broadening that
is $21$~\kms\ and a component from impact broadening that is 9~\kms\
(according to equations 1 and 2 of \citeNP{Garay99} with $n_e = 10^4$
cm$^{-3}$ and $T_e = 10^4$~K).

\begin{figure}
\centering \mbox{\subfigure[Region 1]{\psfig{file=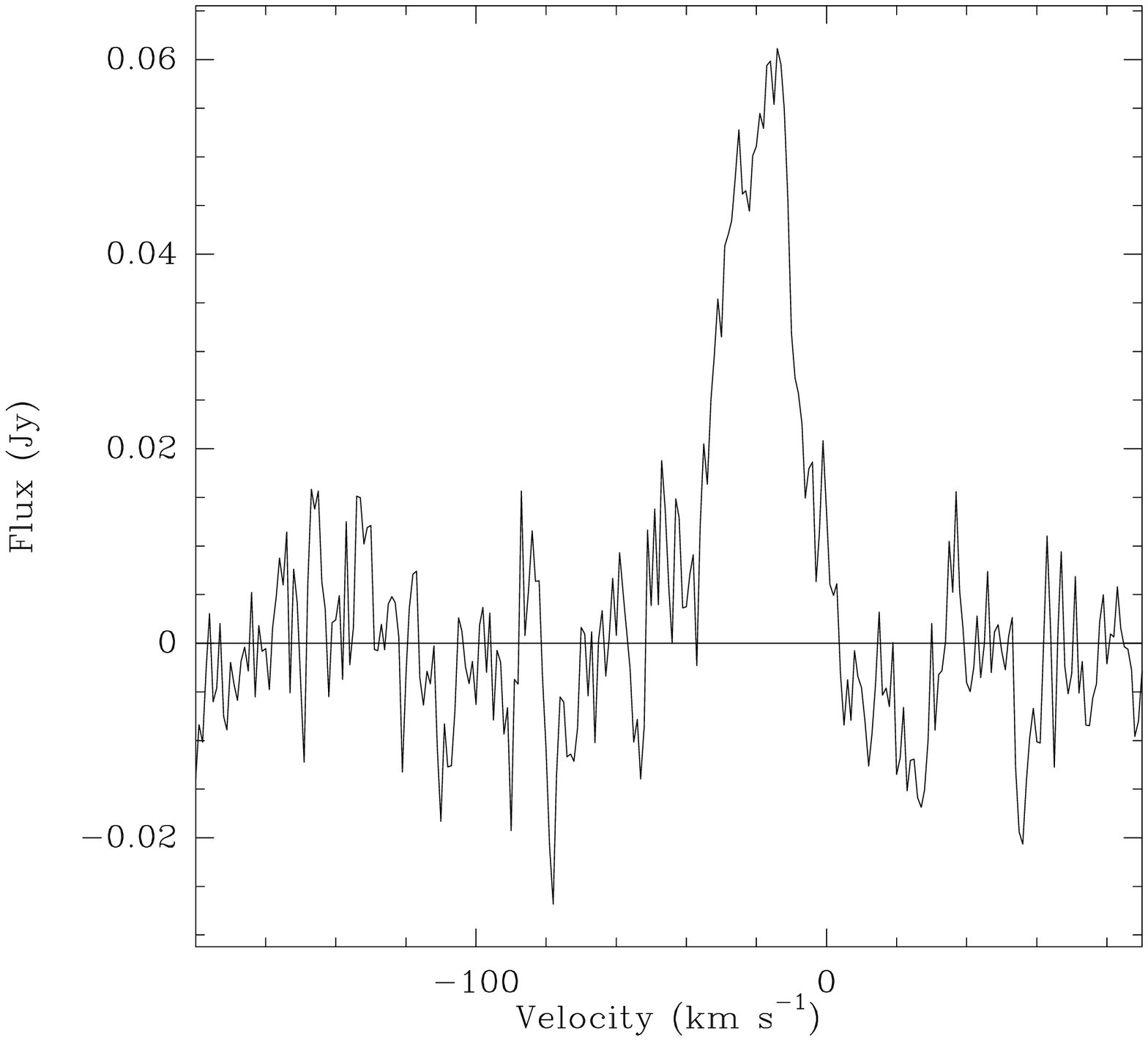,clip=,width=0.2\textwidth,angle=270}}\quad
\subfigure[Region 2]{\psfig{file=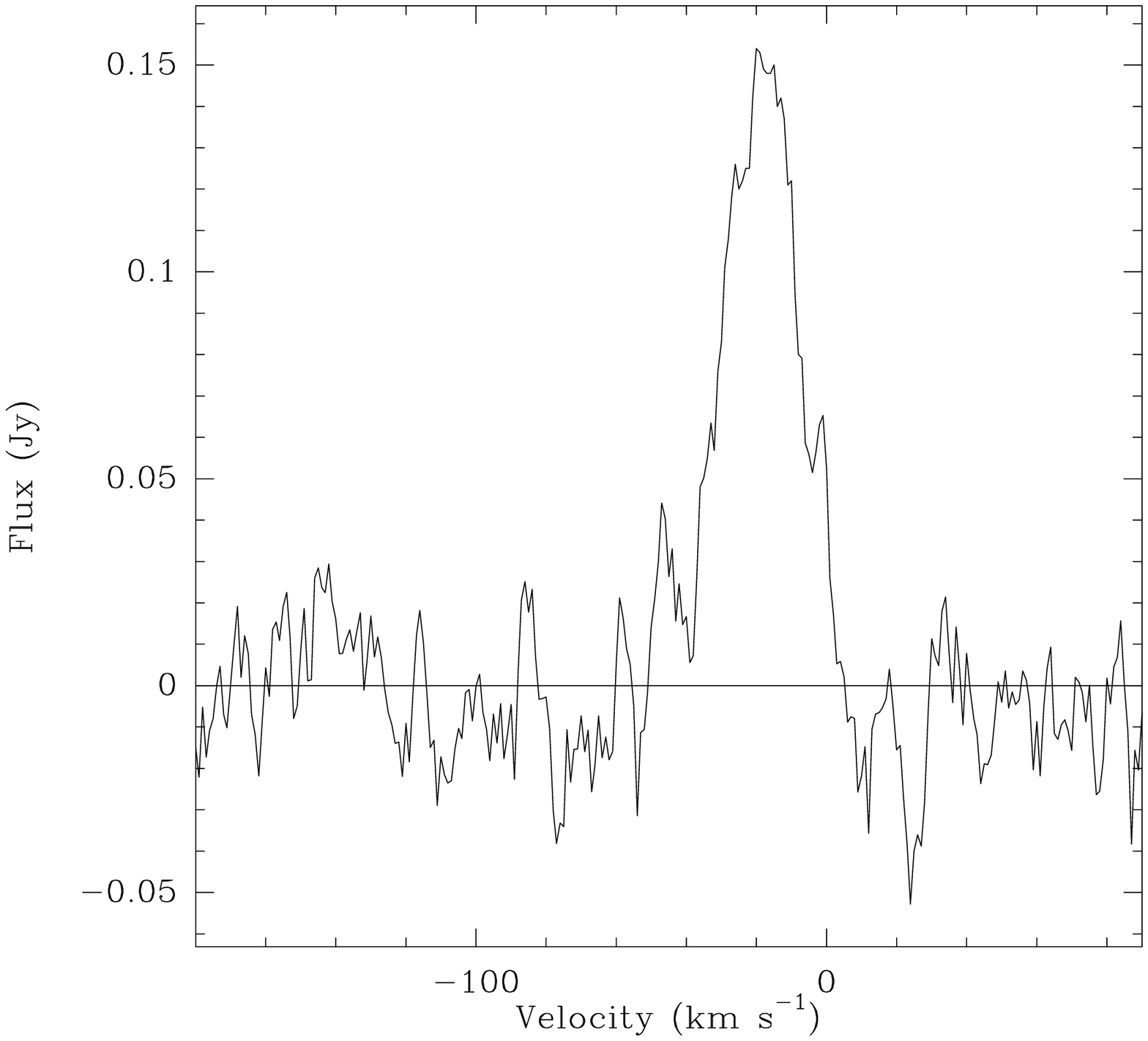,clip=,width=0.2\textwidth,angle=270}}}
\mbox{\subfigure[Region 3]{\psfig{file=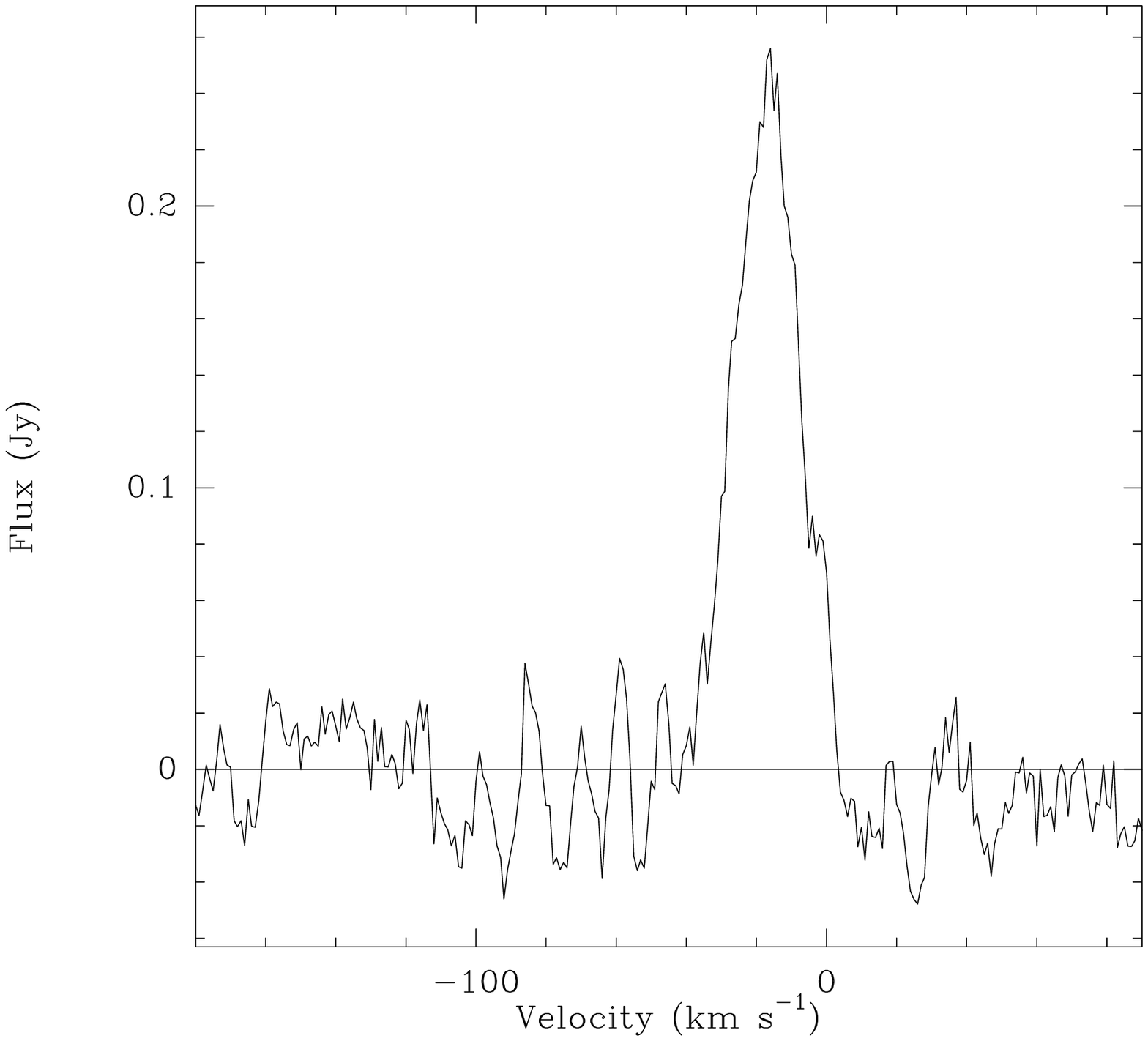,clip=,width=0.2\textwidth,angle=270}}\quad
\subfigure[Region 4]{\psfig{file=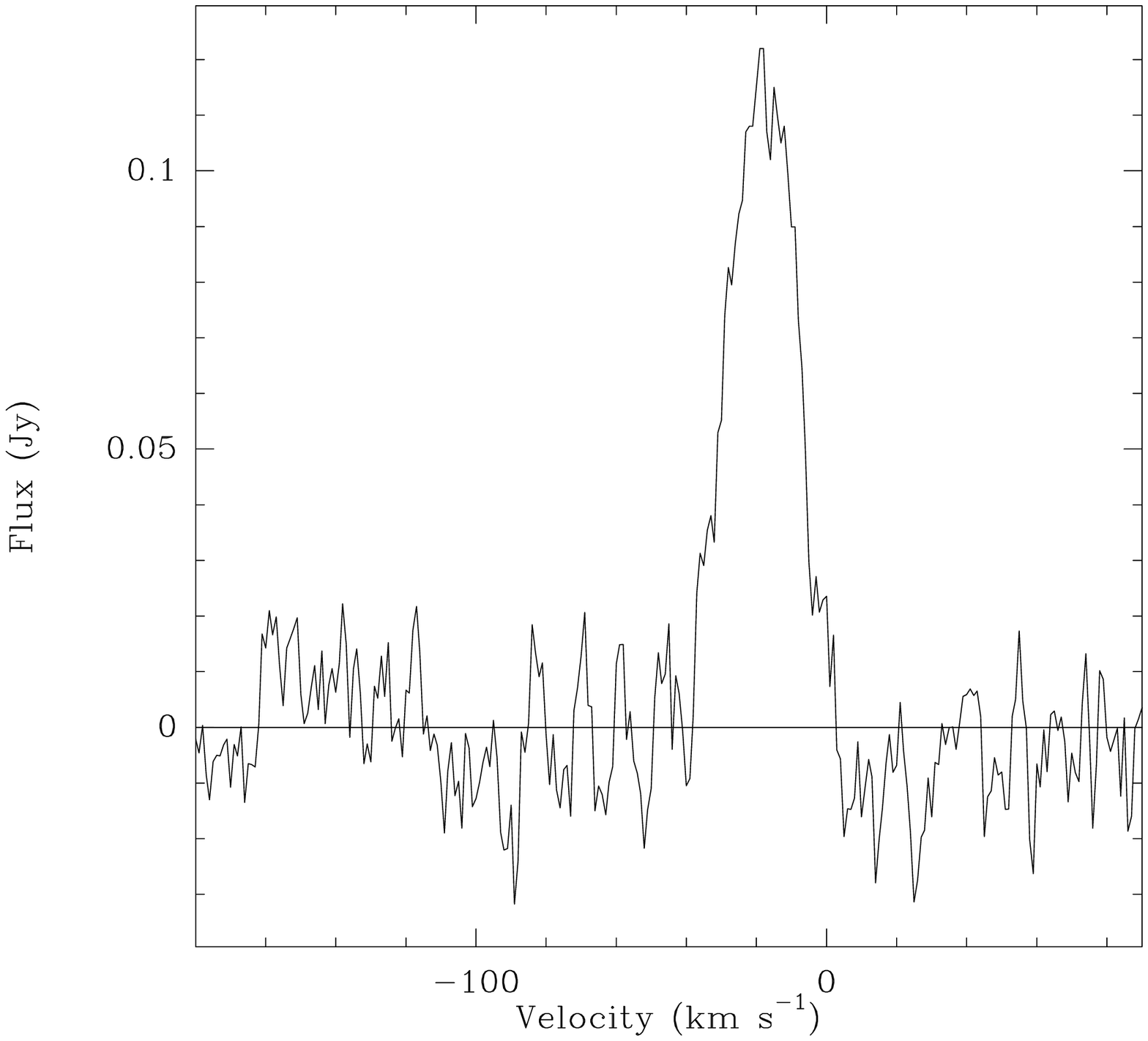,clip=,width=0.2\textwidth,angle=270}}}
\mbox{\subfigure[Region 5]{\psfig{file=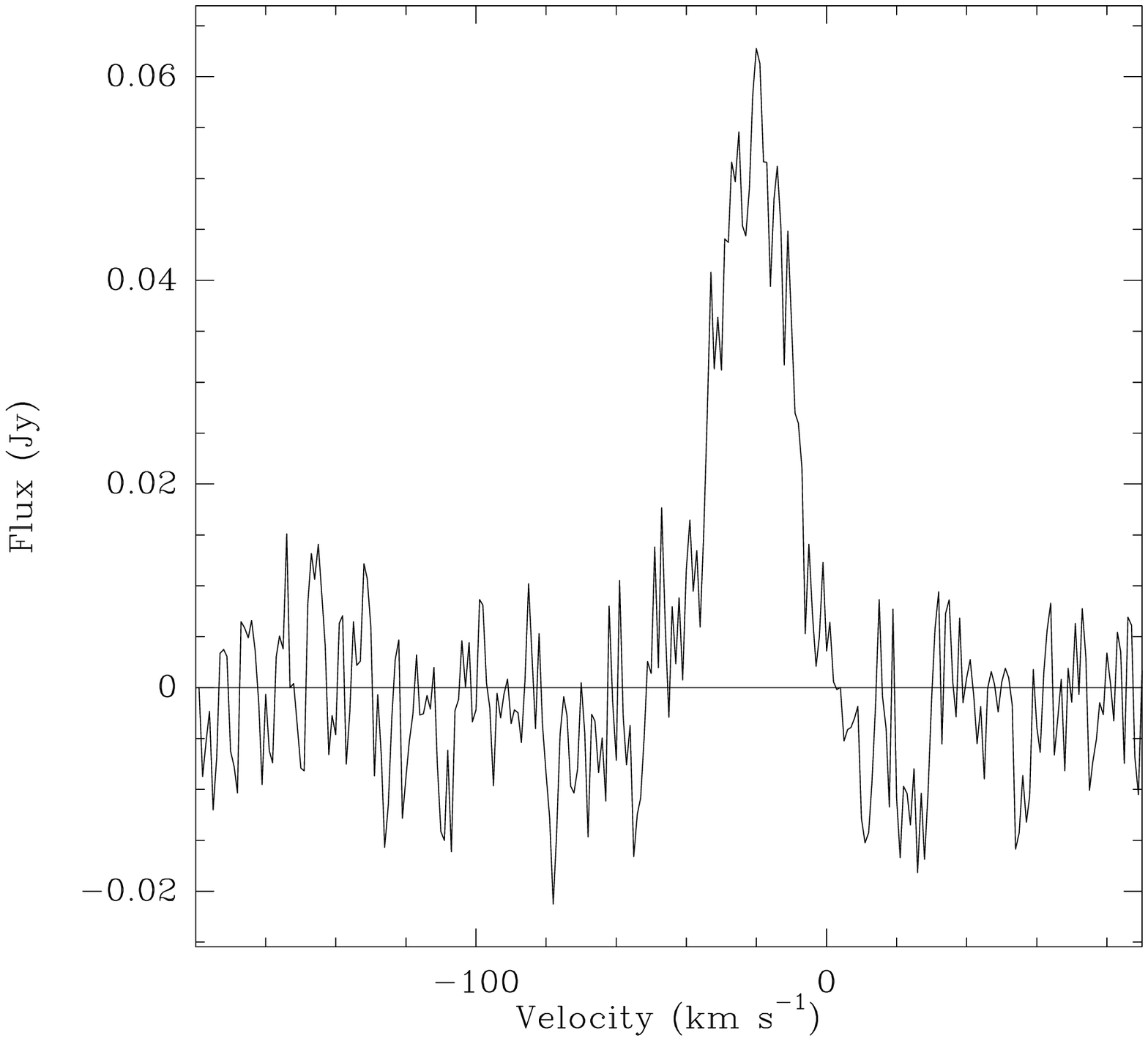,clip=,width=0.2\textwidth,angle=270}}\quad
\subfigure[Total]{\psfig{file=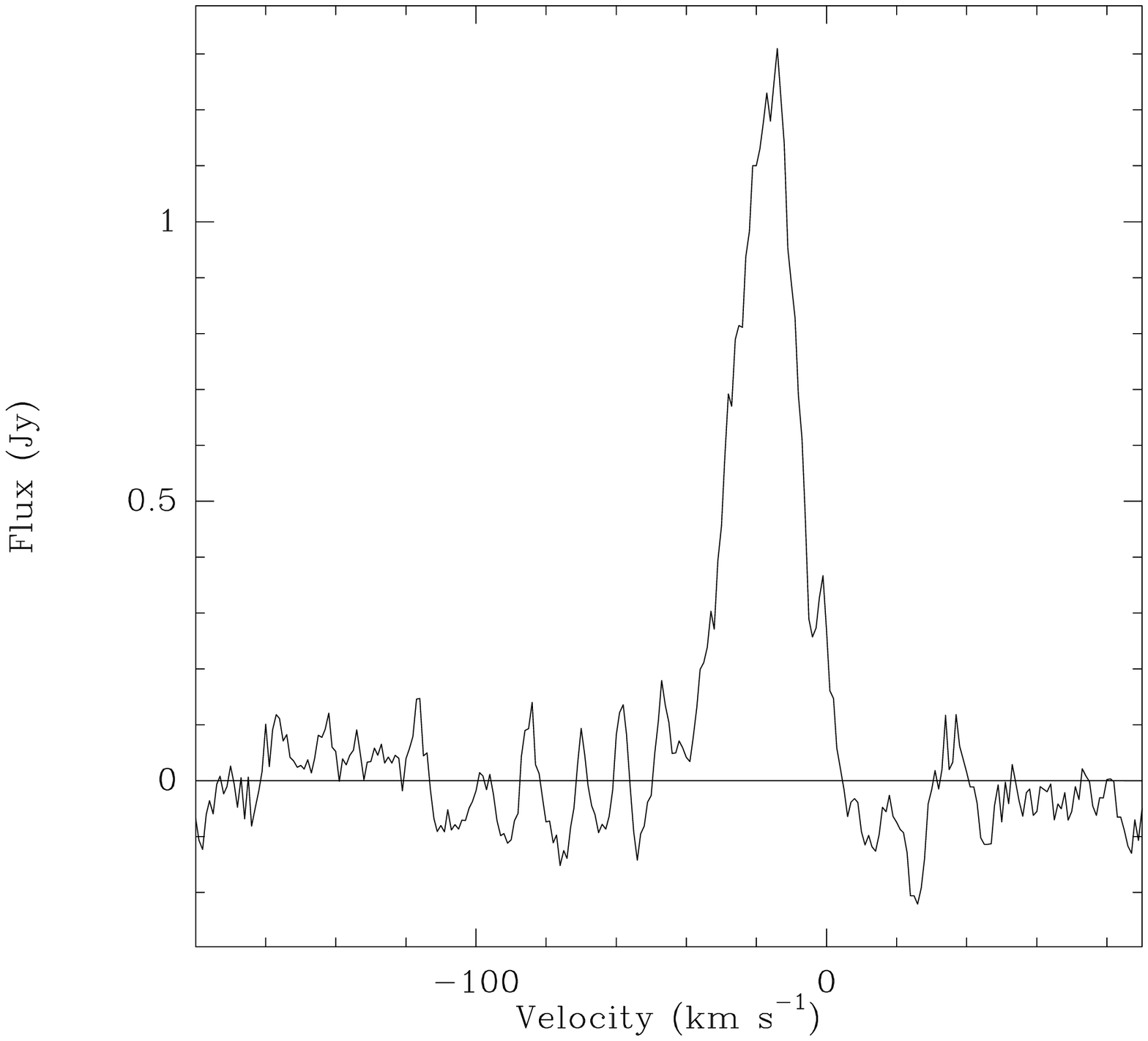,clip=,width=0.2\textwidth,angle=270}}}
\caption{\label{h110_spectra1}Line profiles representing the selected
regions within Car~I (indicated in Fig.~\ref{h110_regions}a).}
\end{figure}

\begin{figure*}
\centering
\renewcommand{\subfigtopskip}{-3pt}
\mbox{\subfigure[Region 1 ]{\psfig{file=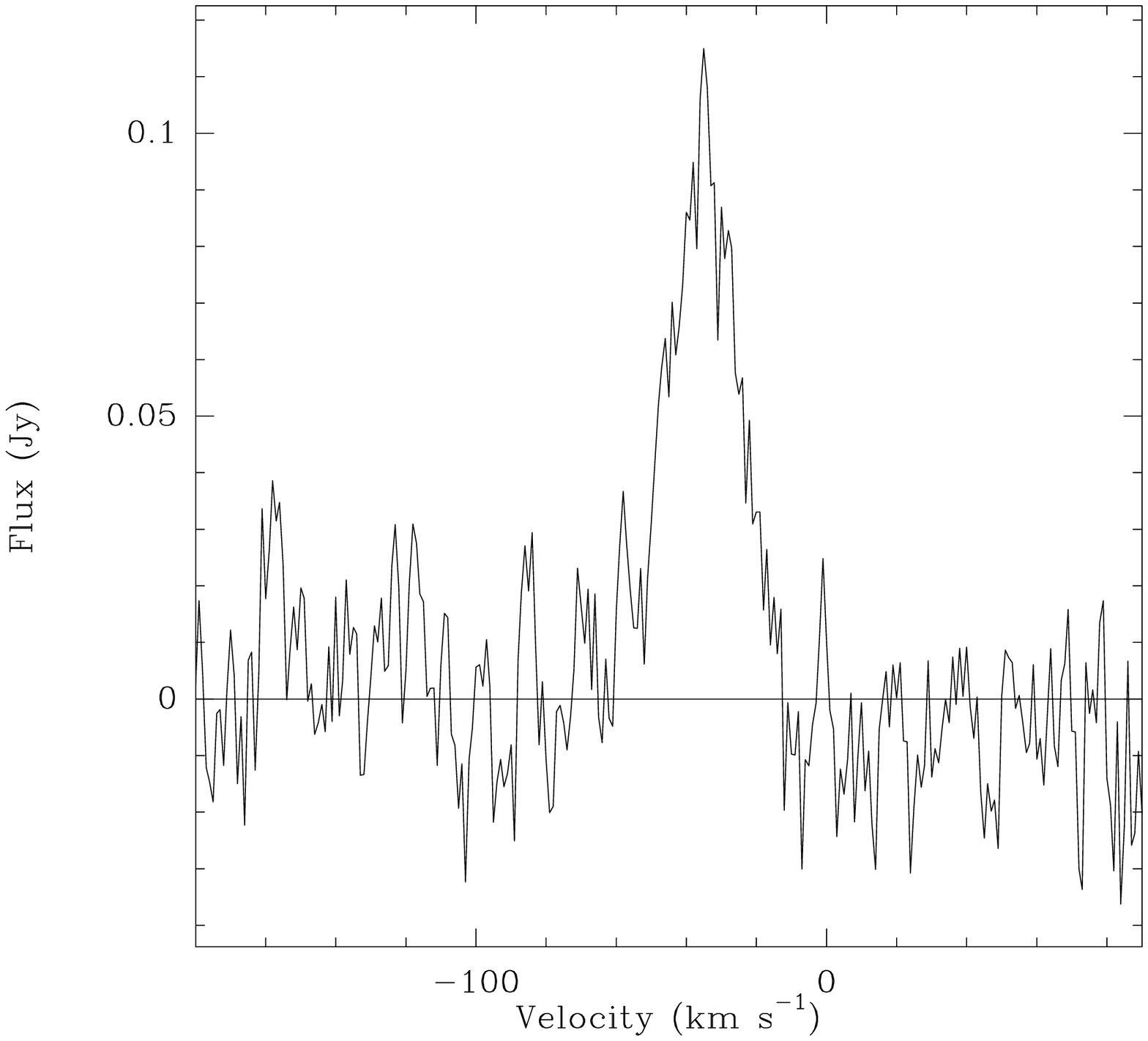
,clip=,width=0.2\textwidth,angle=270}}\quad
\subfigure[Region 2]{\psfig{file=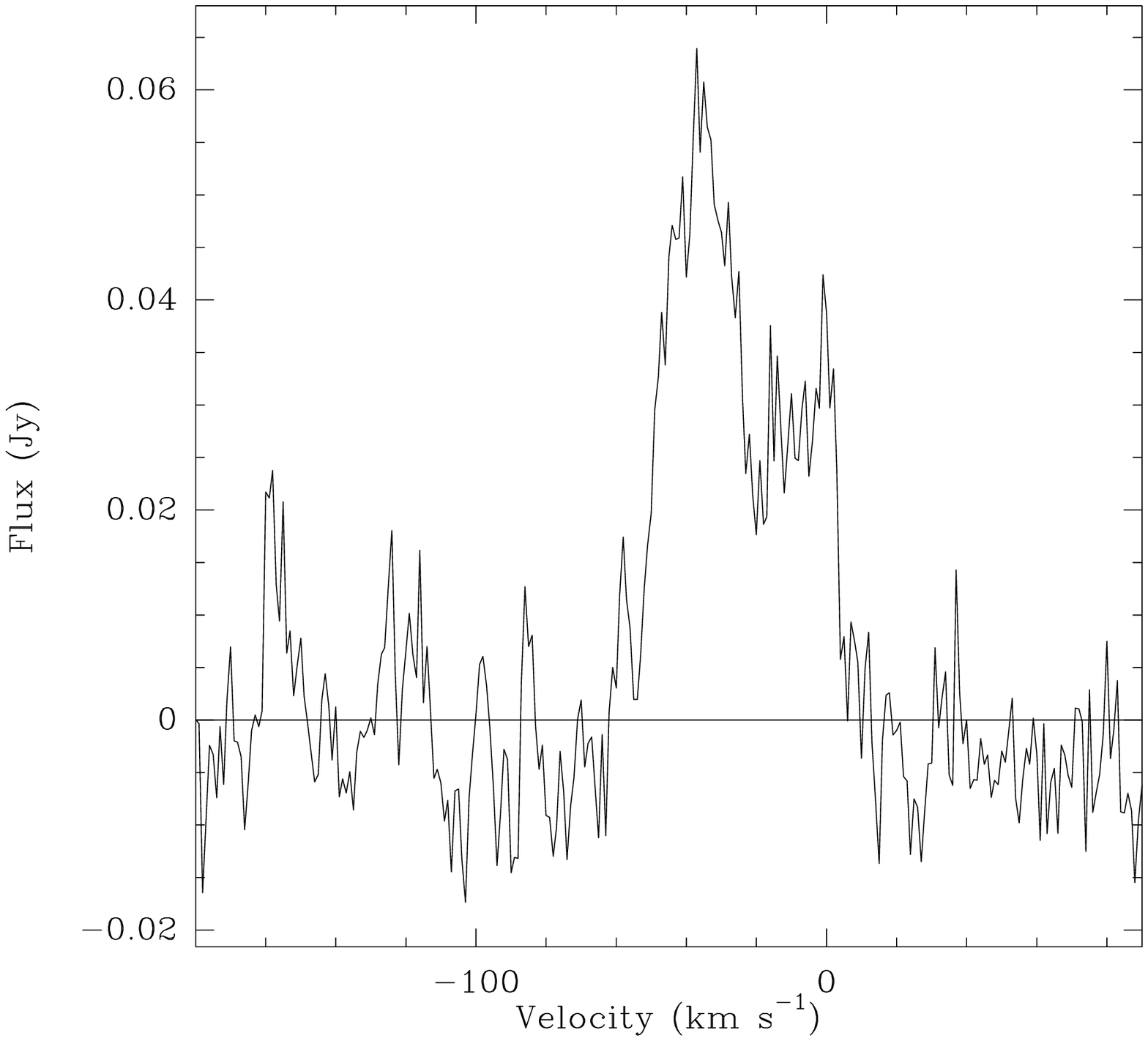,clip=,width=0.2\textwidth,angle=270}}\quad
\subfigure[Region 3]{\psfig{file=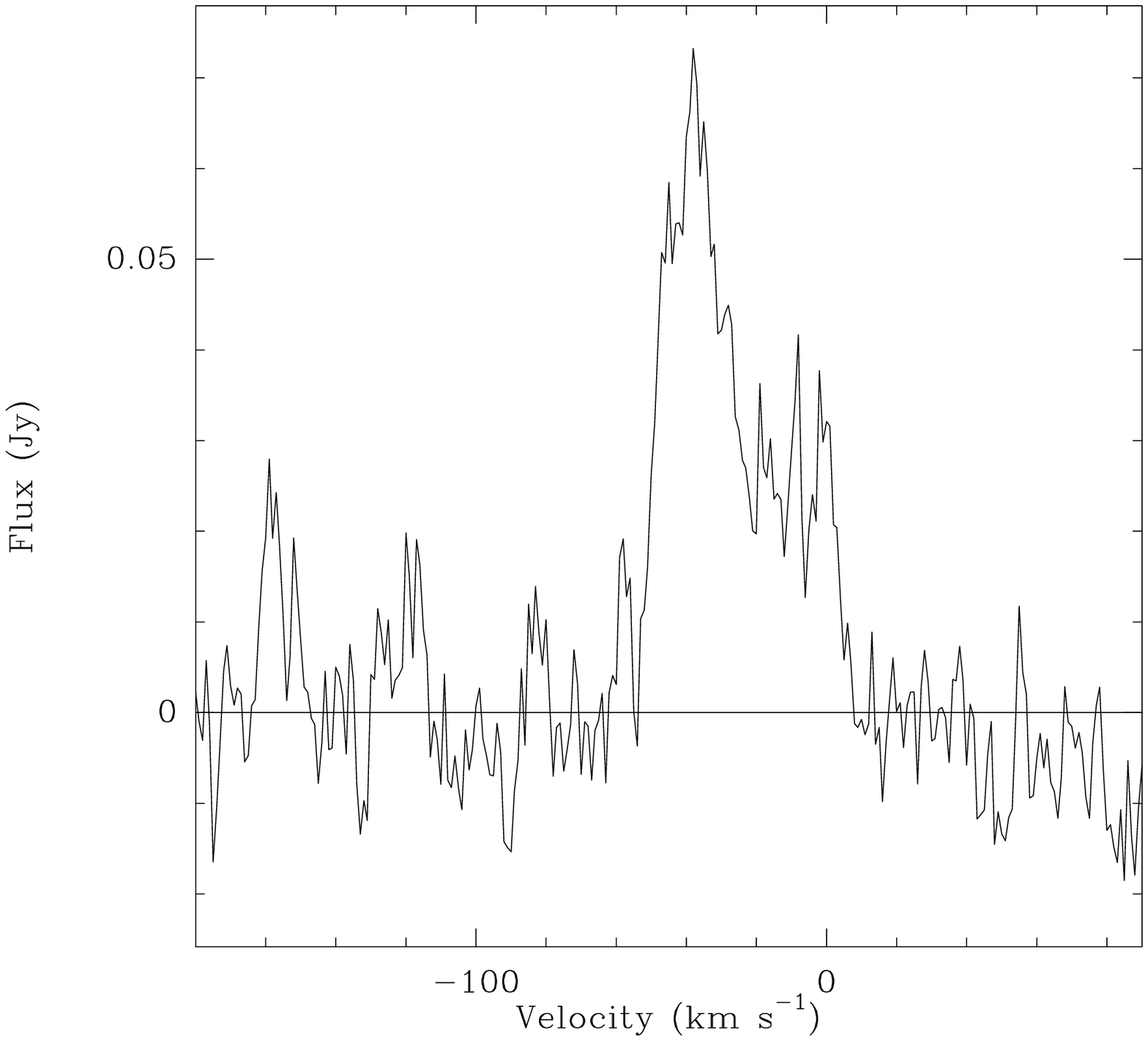,clip=,width=0.2\textwidth,angle=270}}\quad
\subfigure[Region 4]{\psfig{file=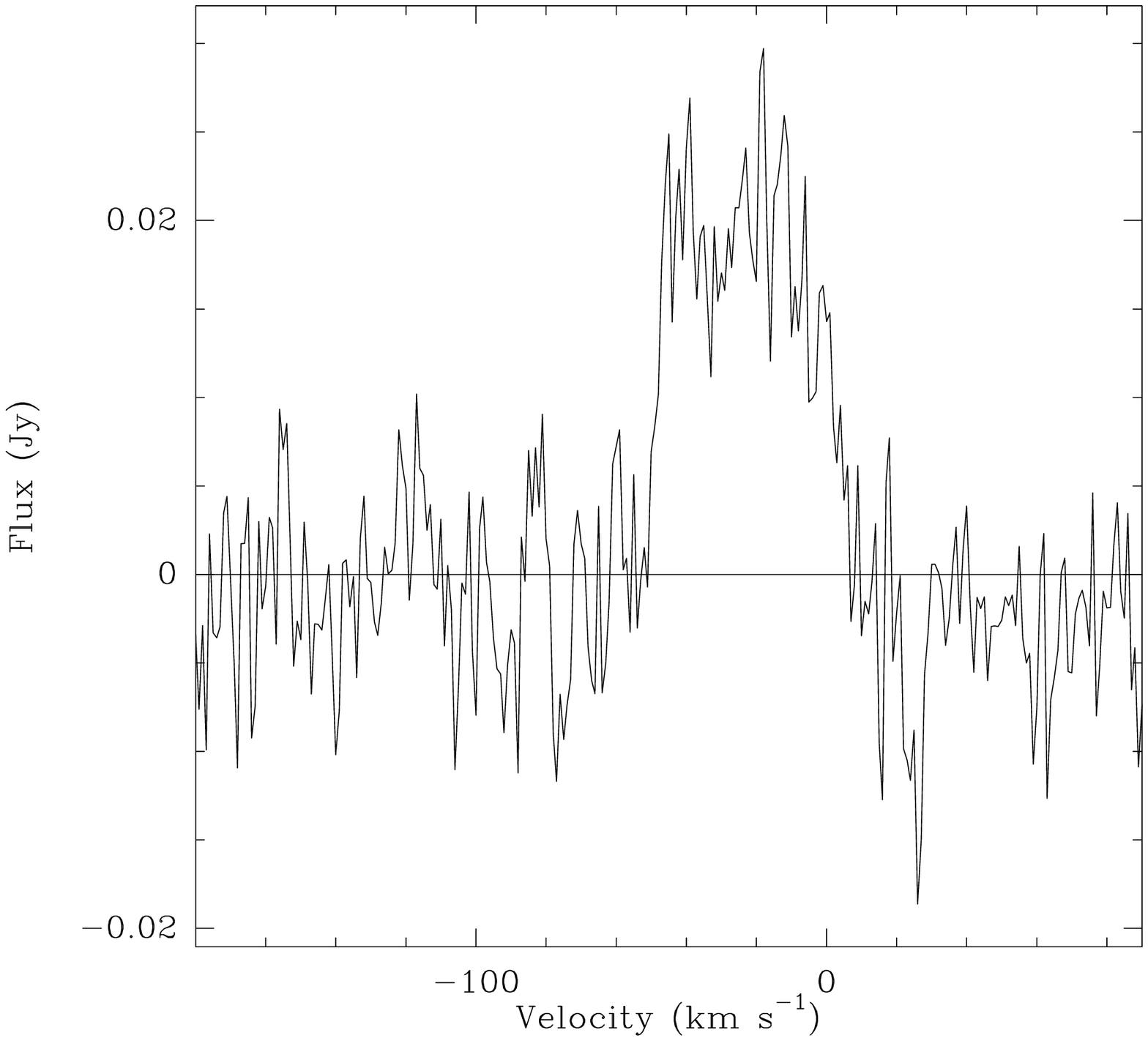,clip=,width=0.2\textwidth,angle=270}}}
\mbox{\subfigure[Region 5]{\psfig{file=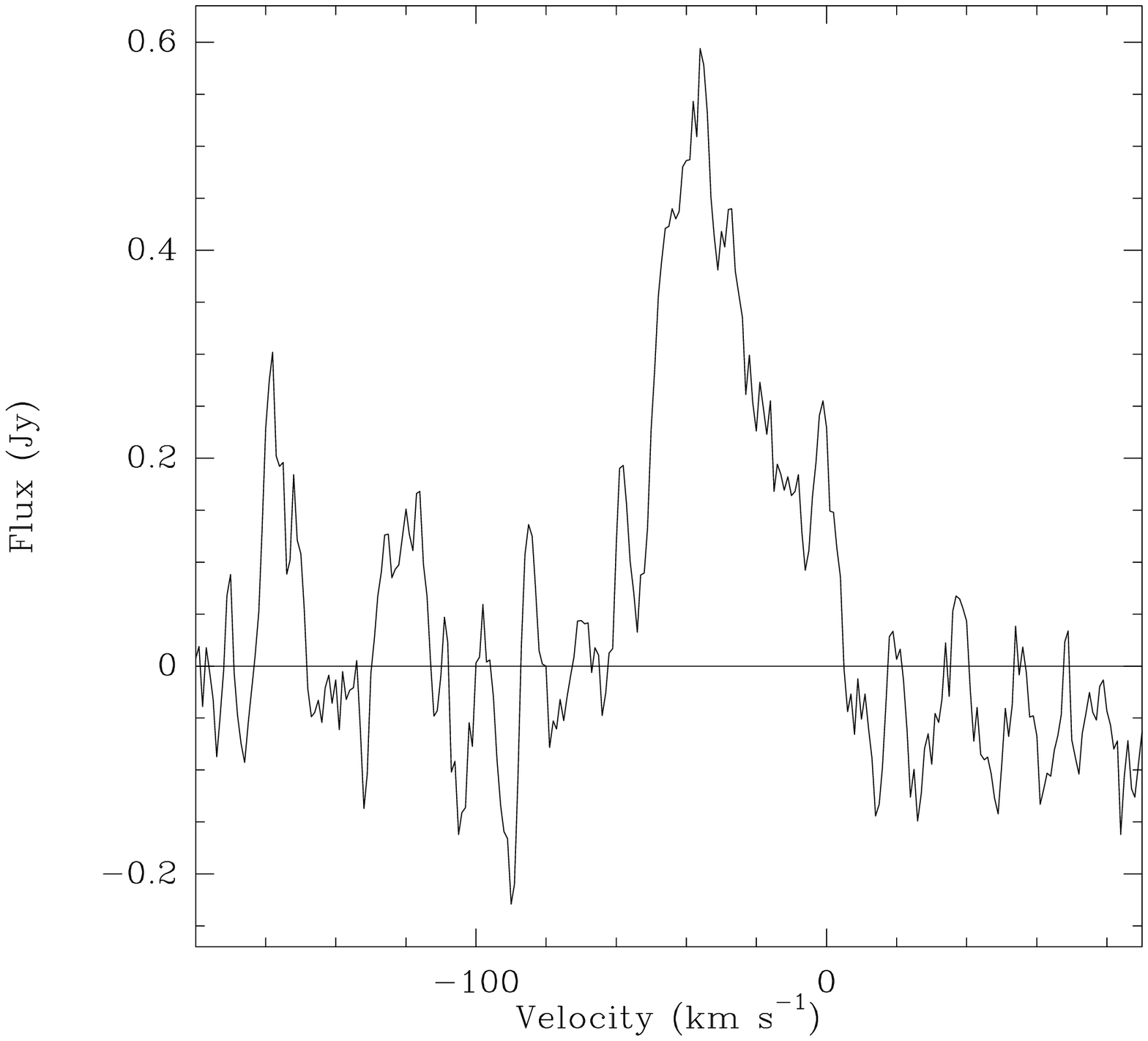,clip=,width=0.2\textwidth,angle=270}}\quad
\subfigure[Region 6]{\psfig{file=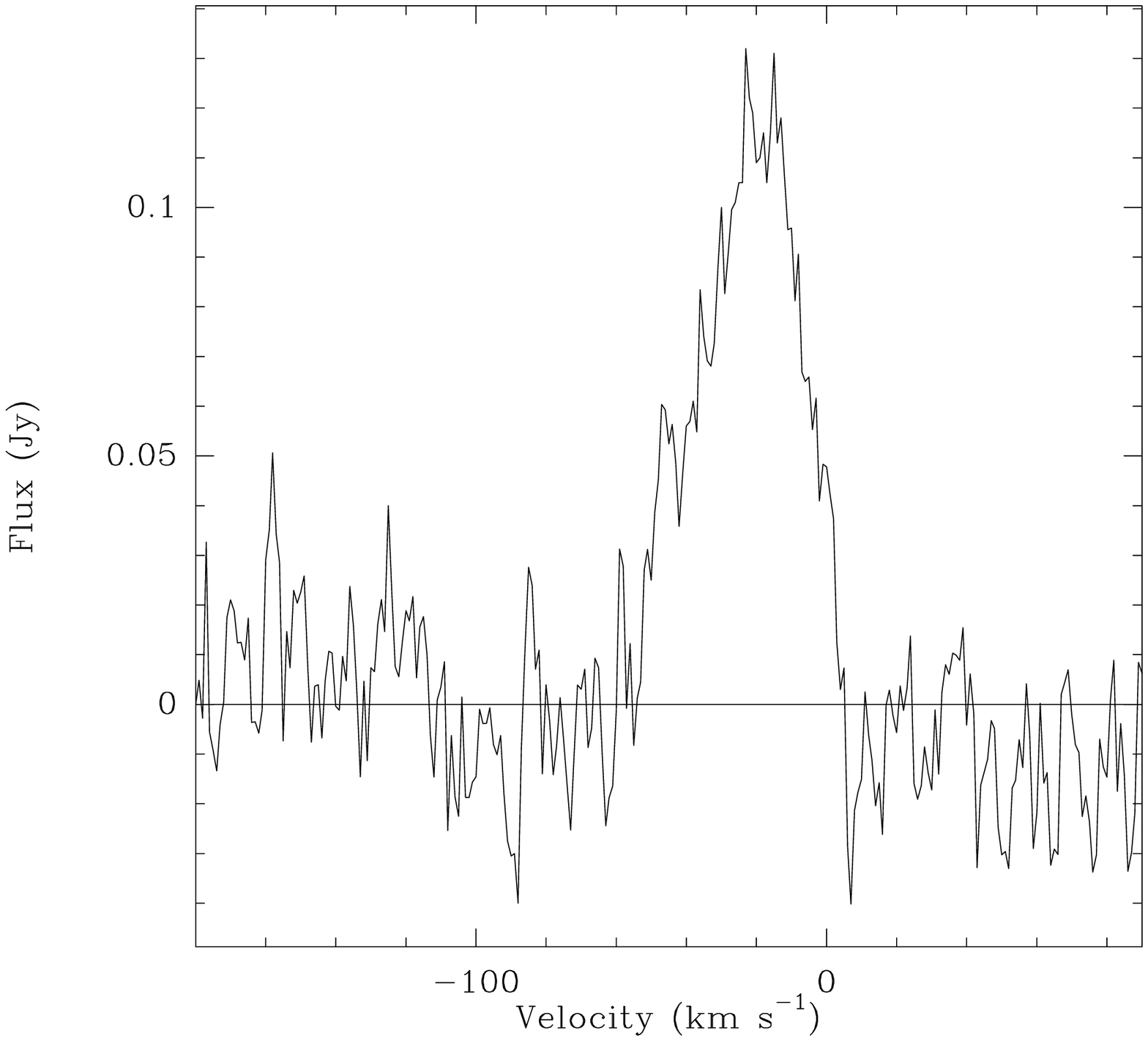,clip=,width=0.2\textwidth,angle=270}}\quad
\subfigure[Region 7]{\psfig{file=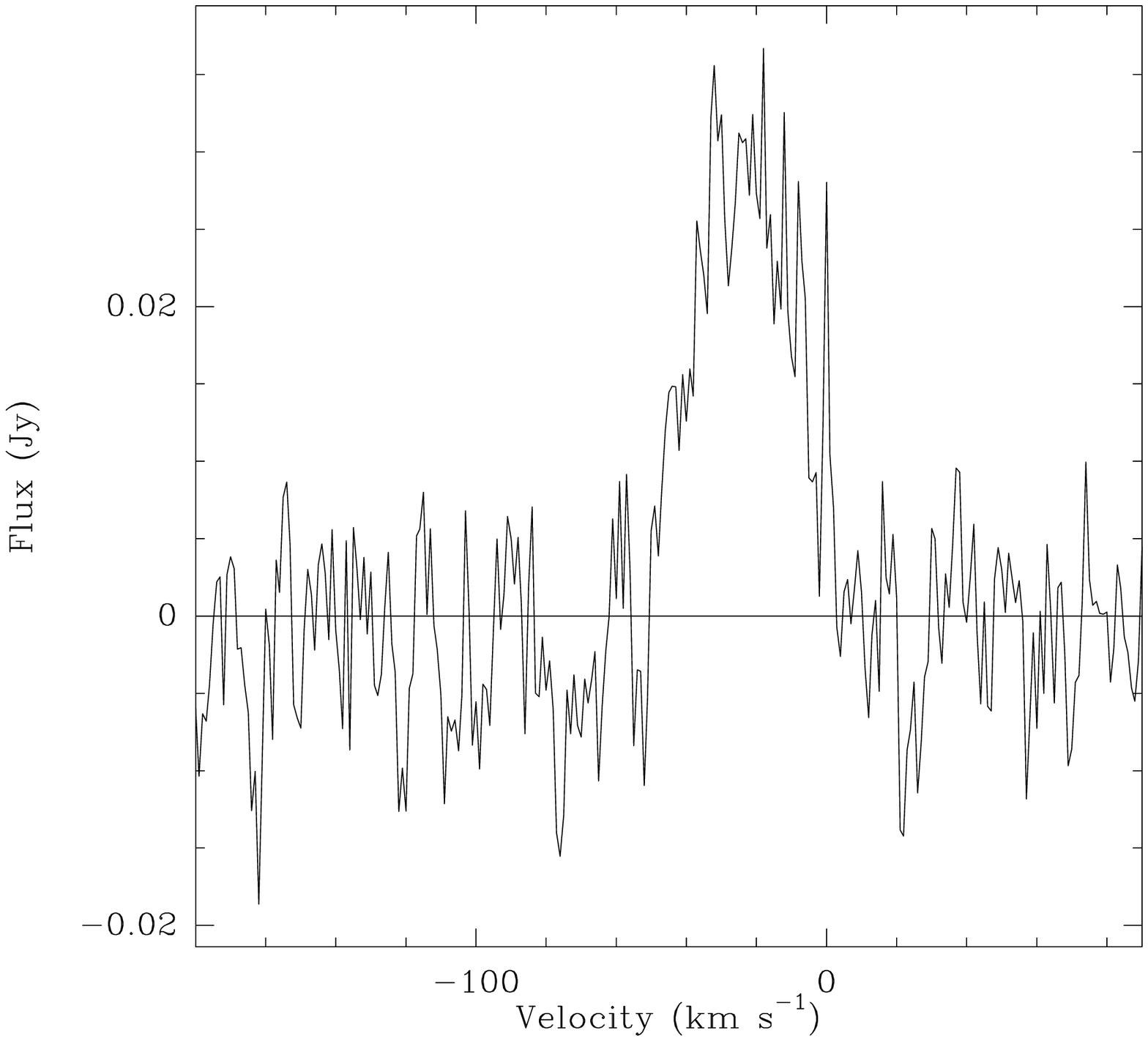,clip=,width=0.2\textwidth,angle=270}}\quad
\subfigure[Total]{\psfig{file=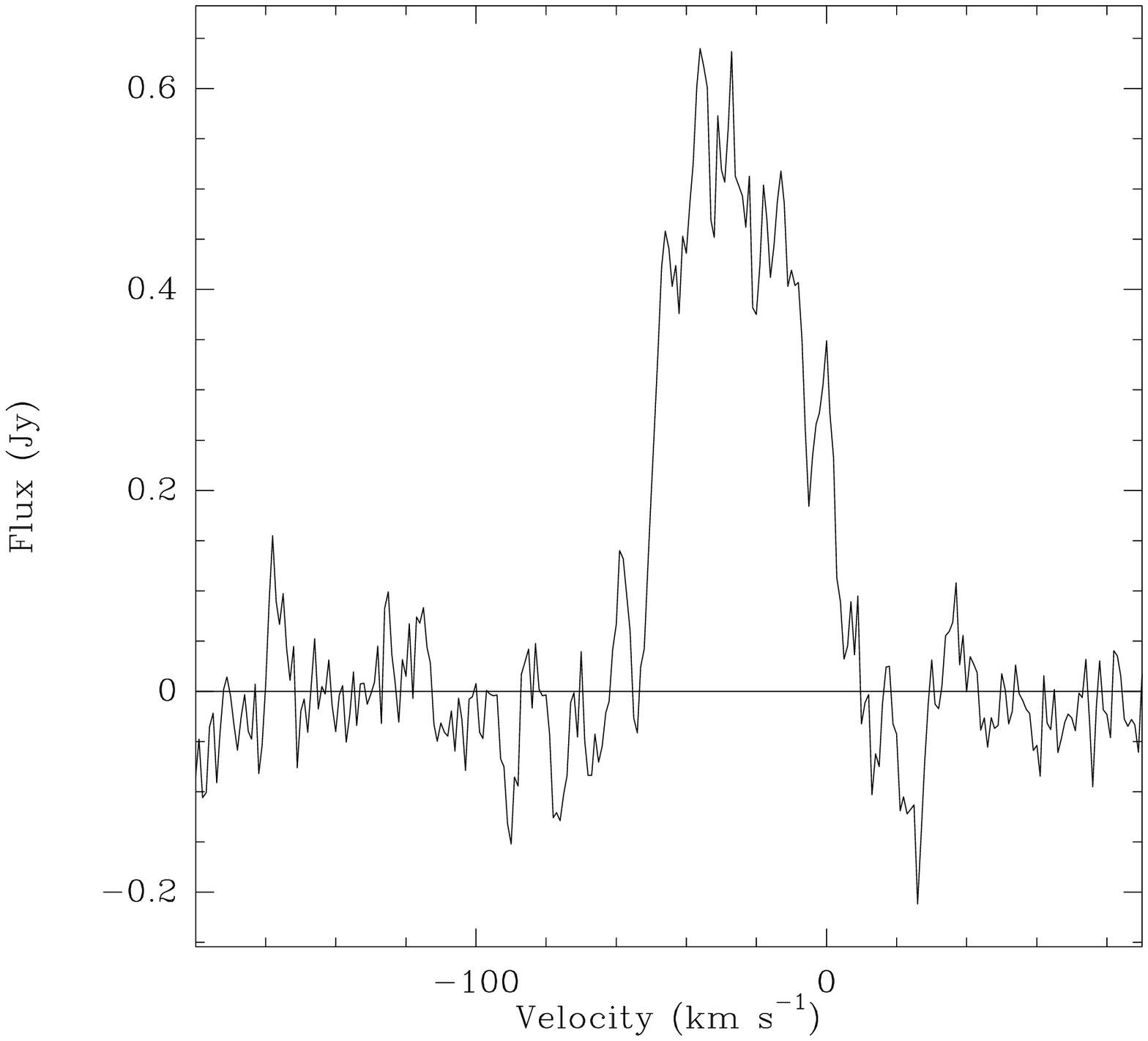,clip=,width=0.2\textwidth,angle=270}}}
\caption{\label{h110_spectra2}Line profiles representing the selected
regions within Car~II (indicated in Fig.~\ref{h110_regions}b).}
\end{figure*}

\begin{table*}
\begin{minipage}{\textwidth}
\caption{\label{h110_params}Results of the Gaussians fits to the
H110$\alpha$ line profiles from the selected regions within Car~I and Car
II. Only fits with a signal-to-noise ratio greater than 3 were
included. Parameters include peak intensity (P), central velocity (V), and
linewidth ($\Delta$V). Also listed for each region are the corresponding
continuum fluxes derived from the natural-weighted 4.8-GHz continuum image
($S_{Total}$ ), the line-to-continuum ratio (L/C) and LTE electron
temperature (T$_e^*$).  Add 122.3~\kms\ and 149.7~\kms\ to the H110$\alpha$
velocities to obtain velocities for He110$\alpha$ and C110$\alpha$
emission, respectively. }
\begin{tabular}{@{}lcccccc}
\hline
Region  & $S_{Total}$   
        & P     
        & V     
        & $\Delta$V       
        & L/C
        & T$_e^*$\\

        & Jy              
        & Jy 
        & \kms         
        & \kms            
        & 
        & K \\
\hline
\it{Car~I}     &       &       &       &       &\\
Total 	        & $25\pm1$      
                & $1.22 \pm 0.02$ 
                & $-17.2 \pm 0.2$
                & $21.0 \pm 0.4$   
                & $0.049 \pm 0.003$      
                & 7 $700 \pm 400$ \\

Region 1        & $0.5 \pm0.1$     
                & $0.057 \pm 0.002$ 
                & $-19.3 \pm 0.5$
                & $23 \pm 1$  
                & $0.13 \pm 0.03$    
                & 3 $100 \pm 700$\\

Region 2        & $1.6\pm0.1$       
                & $0.150 \pm 0.004$          
                & $-18.5 \pm 0.4$
                & $25.3 \pm 0.9$  
                & $0.094 \pm 0.008$
                & 3 $900 \pm 300$     \\

Region 3        & $3.0\pm 0.2$     
                & $0.240 \pm 0.006$
                & $-16.9 \pm 0.3$
                & $21.8 \pm 0.6$
                & $0.080 \pm 0.008$
                & 4 $900 \pm 400$     \\

Region 4        & $1.4\pm 0.1$  
                & $0.120 \pm 0.003$
                & $-18.3 \pm 0.3$
                & $22.2 \pm 0.7$
                & $0.086 \pm 0.008$
                & 4 $300 \pm 400$     \\

Region 5        & $0.4 \pm 0.1$
                & $0.056 \pm 0.002$
                & $-21.0 \pm 0.4$
                & $23.0 \pm 0.9$
                & $0.15 \pm 0.04$
                & 2 $700 \pm 700$      \\
\it{Car~II}     &       &       &       &       &\\

Total           & $24.1 \pm 0.7$       
                & $0.53 \pm 0.04$     
                & $-37 \pm 1$
                & $22 \pm 2$     
                & $0.022 \pm 0.002$     \\
                & 
                & $0.42 \pm0.03$       
                & $-13 \pm 2$      
                & $26 \pm 3$    
                & $0.017 \pm 0.002$     \\
	
		&
		& $0.13 \pm 0.04$
		& $-156.8 \pm 0.6$
		& $4 \pm 1$
		& $0.01 \pm 0.002$			\\

Region 1        & $1.5 \pm 0.2$    
                & $0.095 \pm 0.004$
                & $-35.2 \pm 0.5$
                & $25 \pm 1$
                & $0.06 \pm 0.01$ 
                & 5 $500 \pm 800$    \\

Region 2        & $1.2 \pm 0.1$      
                & $0.056 \pm 0.002$
                & $-35.7 \pm 0.5$
                & $25 \pm 1$       
                & $0.047 \pm 0.005$     \\
                & 
                & $0.032 \pm 0.002$         
                & $-5.0 \pm 0.8$
                & $18 \pm 2$      
                & $0.027 \pm 0.004$     \\

		&
		& $0.020 \pm 0.003$ 
		& $-157.0 \pm 0.8$ 
		& $7 \pm 2$ 
		& $0.02 \pm 0.005$			\\

Region 3        & $1.23 \pm 0.09$      
                & $0.063 \pm 0.004$
                & $-38.7 \pm 0.7$
                & $20 \pm 1$    
                & $0.051 \pm 0.007$     \\
                & 
                & $0.028 \pm 0.002$
                & $-8 \pm 2$
                & $31 \pm 4$        
                & $0.023 \pm 0.003$     \\

		&
		& $0.02 \pm 0.04$ 
		& $-157.1 \pm 0.8$ 
		& $10 \pm 2$
		& $0.02 \pm 0.007$			\\

Region 4        & $0.42 \pm 0.07$    
                & $0.018 \pm 0.003$
                & $-42.3 \pm 0.8$
                & $11 \pm 2$      
                & $0.04 \pm 0.01$     \\
                &       
                & $0.023 \pm 0.001$
                & $-18 \pm 1$
                & $34 \pm 3$      
                & $0.06 \pm 0.01$     \\

Region 5        & $8.4 \pm 0.5$     
                & $0.52 \pm 0.08$
                & $-37 \pm 2$
                & $25 \pm 4$  
                & $0.06 \pm 0.01 $  \\
                & 
                & $0.17 \pm 0.03$
                & $-7 \pm 2$
                & $26 \pm 4   $   
                & $0.020 \pm 0.005$     \\
	
		&
		& $0.25 \pm 0.03$ 
		& $-156.9 \pm 0.5$ 
		& $7 \pm 1$
		& $0.03 \pm 0.005$			\\

		&
		& $0.12 \pm 0.03$ 
		& $-120 \pm 1$ 
		& $11 \pm 2$
		& $0.01 \pm 0.005$			\\

Region 6        & $2.2 \pm 0.5$ 
                & $0.04 \pm 0.01$
                & $-42 \pm 2$
                & $18 \pm 3$
                & $0.018 \pm 0.004 $   \\
                &
                & $0.122 \pm 0.008$
                & $-11 \pm 1$
                & $26 \pm 3$
                & $0.06 \pm 0.02$\\

Region 7        & $0.32 \pm 0.2 $
                & $0.02 \pm 0.01$
                & $-34 \pm 2$
                & $20 \pm 3    $  
                & $0.06 \pm 0.07$\\
                &
                & $0.024 \pm 0.009$
                & $-16 \pm 2$
                & $26 \pm 3    $ 
                & $0.08 \pm 0.08$       \\
\hline 
\end{tabular}
\end{minipage} 
\end{table*}

The line profiles from the regions in Car~I are shown in
Fig.~\ref{h110_spectra1}. Each profile contains one component that is
centred between velocities of $-$17 and \mbox{$-21$~\kms}, with no apparent
spatial trend from region to region.  The linewidth values are between 21
and 25~\kms, consistent with thermal motions. The previously detected
H109$\alpha$ emission profile centred on Car~I has one single
component, centred at a velocity of $-17$~\kms\ \cite{Huchtmeier75}. This is
consistent with the H110$\alpha$ value of \mbox{$-17.2 \pm 0.2$}, taken
from the total integrated emission line profile.

The line profiles from Car~II are shown in Fig.~\ref{h110_spectra2}. All
profiles show two peaks, with the exception of the profile from Region
1. This region is located at the southern edge of Car~II-W and represents
the only region in Car~II where the profiles do not exhibit two
well-defined peaks. The single peak is centred at a velocity of $-35.2 \pm
0.5$~\kms\ with a linewidth of $25 \pm 1$~\kms. For all the other regions,
the central velocity of one component is between $-$42 and $-$33~\kms\ and
the other between $-$18 and 4~\kms. There appears an overall trend in the
velocity separation of the component pairs: in the southern regions
(Regions 2, 3 and 5) the velocity separation is \mbox{$\sim30$~\kms}; in
the central east (Regions 4 and 6) the velocity separation is \mbox{$\sim
24$~\kms}; and for the north (Region 7) the velocity separation is
\mbox{$\sim 17$~\kms}. The line-to-continuum values for each component vary
between 0.02 and 0.08, with no clear spatial trend for the relative values
of each component pair.

Values for the linewidths of each component in Car~II vary between 11 and
34~\kms, with an average value of 24~\kms. The two extremities in this
range are attributed to the profile from Region 4 and largely differ from
the thermal linewidth value of \mbox{$\sim 22$~\kms}. Narrow linewidths
of \mbox{$ \leq 10$~\kms} are frequently detected in recombination-line
studies and are understood to arise from PDRs
(e.g.~\citeNP{Roelfsema91}). This may be the case for the narrow
component, at \mbox{$\sim -42$~\kms}, from Region 4. The broad component
arising from the same region at \mbox{$\sim -18$~\kms} may be the result
of broadening mechanisms such as electron impacts and non-thermal
broadening. In this case, a linewidth of $34$~\kms\ corresponds to a
non-thermal component of $ \sim 25$~\kms\ (using equation 3 of
\citeNP{Garay99} with \mbox{$n_e = 10^4$ cm$^{-3}$} and \mbox{$T_e =
10^4$~K}). All the line profiles are likely to contain contributions from
non-thermal motions, and possibly from PDRs. It is not clear why Region 4,
which represents a small region in Car~II-N, is the most affected.

Individual data points of H$\alpha$ and [N{\sc ii}] emission taken by
\citeN{Deharveng75} also show strong line splitting in the vicinity of
Car~II.  The H109$\alpha$ emission profile centred on Car~II has two
components with central velocities of $-$33
and $-$8~\kms\ \cite{Huchtmeier75}. The velocity separation is consistent
with the H110$\alpha$ value of 24~\kms\ whereas the central velocities are
offset by 3~\kms. This offset is likely to be caused by the different
regions over which each profile was obtained. It has been noted in W3(OH)
however, that the centre velocity of a radio recombination line increases
with the frequency of the transition. This has been explained by optical
depth effects in a dense and expanding ionized gas shell (see
\citeNP{Garay90} and references therein).

An interesting feature of the Car~II emission profiles, is the presence of
two faint and narrow components centred near velocities of $-$120 and
$-$157~\kms\ which are discussed further in Section~\ref{HeC}.

The parameters of the Gaussian fits to the emission line profiles listed in
Table~\ref{h110_params} have been used to obtain values for the LTE
electron temperature ($T_e ^{\star}$) according to equation~22 of
\citeN{Roelfsema92}. In the case of Car~II, only a single measurement for
$T_e ^{\star}$ was obtained (from the single-peaked profile of Region 1). A
typical value of 0.1 was taken for the fractional helium abundance
($Y^+$). The uncertainties in the measured line-to-continuum values
(primarily caused by the negative continuum emission `bowl') lead to an
uncertainty in $T_e ^{\star}$ of typically \mbox{25~per~cent}. Values range
between 2700~K and 7700~K and are consistent with the single value of 6$600
\pm 500$~K derived from the H109$\alpha$ profile \cite{Huchtmeier75}.

\section{Discussion}
\subsection{\label{carI}Car~I}

The arc-shaped emission of Car~I-E coincides remarkably well with a
semicircular bright-rimmed optical feature. Such a coincidence was also
noted by \citeN{Whiteoak94}. It was interpreted as a ionization front
bordering dense molecular gas associated with the dust lane. At the edge of
the dust lane there is a strong CO emission concentration between $-$35 and
$-$21~\kms\ which is likely to be a dense clump in the front face of the
molecular cloud (see \citeNP{Brooks98}). It is adjacent to both Car~I-E and
the bright-rimmed optical feature. The way in which Car~I-E curves around
the dense molecular clump is suggestive of a strong interaction taking
place between the molecular and ionized gas. Car~I-W is located further
into the dust lane and is adjacent to another strong CO concentration
between $-$20 and 0 \kms. This may be a dense clump at the back face of the
molecular cloud and which Car~I-W curves around.

The results presented here clearly support the view that Car~I-E and Car
I-W are strong ionization fronts which are adjacent to dense parts of a
molecular cloud. Their curved morphology supports the currently held view
that these ionization fronts are from Tr~14. Furthermore, by combining the
results from a photometric study of 475 stellar members of Tr~14 by
\cite{Vazquez96} with the stellar energy distributions modeled by
\citeN{Kurucz97}, the total ionizing luminosity of Tr~14 is estimated to be
$\sim 59 \times 10^5$~\Lsun (see \citeNP{Brooks00t}). The bulk of this
luminosity is centered at the core of Tr~14, approximately 3 arcmin from
the position of Car~I-E and 5 arcmin from the position of Car~I-W. Assuming
zero projection to Tr~14 and adopting 1.5 arcmin in extent for both Car~I-E
and Car~I-W, yields values of $2 \times 10^{48}$ photon s$^{-1}$ and $0.8
\times 10^{48}$ photon s$^{-1}$ arriving at Car~I-E and Car~I-W,
respectively. These values are of the order of those derived from the
measured 4.8-GHz continuum fluxes in Table~\ref{h110_table1}.

The two detected compact \HII\ regions are located towards the dust lane
(and associated molecular cloud). If they are early type BO and O9.5 stars,
as their radio fluxes suggest (see Table~\ref{contparam}), then this is the
first evidence of ongoing star formation within the northern cloud of
Carina Nebula.  These sources are offset from the main Tr~14 cluster and
are in close proximity to the interface between the ionization fronts and
the molecular cloud. They may be examples of shock induced star formation,
first discussed by \citeN{Elmegreen77}. Infrared studies would be useful to
establish if indeed the two detected compact \HII\ regions are young
embedded stars.

The molecular gas appears to wrap around most of Car~I suggesting that
Car~I may be carving out an ionized gas cavity at the edge of the GMC. The
H110$\alpha$ line data described here do not exhibit the characteristic
profiles of expansion (with two peaks). However, this does not necessarily
rule out expansion. Instead it could imply that the expansion velocity is
smaller than the sound speed of the ionized gas. In this case the linewidth
would be dominated by the thermal component. Studies have shown that
observed linewidths are very broad for ultra-compact \HII\ and quite narrow
for more extended ones (see \citeNP{Garay99}). These results imply that the
contribution from turbulent motions decays with time. Such behaviour is to
be expected for the simple evolution of an expanding \HII\ region. Under
this premise, the single profiles of Car~I are consistent with an age
of $\sim 10^6$~yr.

\subsection{Car~II}

Contrary to what is seen in Car~I, the strongest continuum emission
features have bright optical emission counterparts. This was noted
previously in other studies (\citeNP{Whiteoak94} and
\citeNP{Retallack83}). However, the higher-resolution data shown here
illustrate that this similarity extends to the scale of the small emission
knots and filaments, and in particular to the narrow ring feature,
Car~II-N.

The origin of Car~II-N remains unknown. \citeN{Retallack83} have suggested
that it is part of an expanding shell, powered by a star or stars within
it. This idea was based on a bright peak of far-infrared emission which is
located within the ring \cite{Harvey79}. However, the study by
\citeN{Deharveng75} reveal strong [OIII] emission outside the ring while
strong [NII] emission is associated with the interior. These results are
consistent with the scenario proposed by \citeN{Harvey79} in which Car~II
is excited externally by the numerous nearby stars of Tr~16, and the ring
is an ionization front that is enveloping warm molecular gas. A small
molecular clump (Clump 5) has been detected by \citeN{Cox951} in the
vicinity of the emission ring. However, this clump is likely to be situated
in front of the nebula, and removed from the ionized gas \cite{Brooks00}.

The bulk of the ionizing luminosity of Tr~16 can be attributed to
$\eta$~Car alone (see \citeNP{Brooks00}). Using a stellar radius of $6
\times 10^{12}$~cm \cite{Davidson97} and an effective temperature of $\sim
50$~000~K \cite{Kurucz97} yields a value of $7 \times 10^{50}$
photons s$^{-1}$ leaving $\eta$~Car. Assuming zero projection and adopting
a separation of 4.5 arcmin and an extent of 1.5 arcmin for Car~II-N, the
estimated number of ionizing photons arriving at this region from
$\eta$~Car is $\sim 5 \times 10^{48}$ photons s$^{-1}$. This value is
sufficient to explain the value determined from the 4.8-GHz continuum
measurement (see Table~\ref{h110_cont_l}), supporting the idea that
Car~II-N is an ionization front originating from $\eta$~Car.

The origin of the continuum emission bar (Car~II-E and Car~II-W) remains a
puzzle. Its sharp southern boundary, which faces $\eta$~Car, may be a
result of depleted gas. For its 4.8-GHz continuum flux to be consistent
with ionization by $\eta$~Car, a projection angle of $\sim 80$~deg needs to
be invoked.

A number of position-velocity traces were taken across different parts of
Car II. Combined with the previously discussed channel maps and line
profiles, the velocity trends are consistent with two main emission
components at red-shifted ($-$5 to $-18$~\kms) and blue-shifted ($-34$ to
$-42$~\kms) velocities. These two components could be indicative of
separate \HII\ regions at different distances along the line of
sight. However, in this chance coincidence, we would expect the line
separation to remain constant across the source. Our results however, show
that this is not the case and a maximum line separation of 30~\kms\ is
evident towards the central part of Car II. \citeN{Deharveng75} suggest an
expanding shell centred on $\eta$~Car with an expansion velocity greater
that 25~\kms. Although our results do not rule out the
possibility of expansion from $\eta$~Car, they are more consistent with
expansion from the location defined by Region 2,  with an expansion
velocity of 15~\kms.

Given the extreme conditions that Car~II is exposed to as a result of its
close proximity to both Tr~14 and Tr~16, a more favorable scenario could
be that there are two (or more) expanding shells in the vicinity of Car~II, of
which the two main H110$\alpha$ velocity components are fragments.

\subsubsection{Additional Components}

\label{HeC}

A number of the H110$\alpha$ emission profiles towards Car~II contain faint
components at velocities close to $-$157~\kms\ and $-$120~\kms\ (see
Table~\ref{h110_params}). Region 5 in Car~II defines the area over which
the additional components are detectable. This region corresponds to the
southern part of Car~II-E and Car~II-W (with main H110$\alpha$ components
at $-$7 and $-$37 \kms). The narrow linewidth of the component at
$-$157~\kms\ is typical of that for C110$\alpha$ emission. Carbon emission
is commonly detected towards bright \HII\ regions and is thought to arise
from PDRs (e.g. \citeNP{Garay98} and \citeNP{Kantharia98}). In this case
the central velocity of the component translates to $-$7 \kms. The broader
linewidth of the component at $-$120~\kms\ is consistent with He
110$\alpha$ emission and translates to a velocity of 2 \kms. However, this
velocity is considerably offset from the main hydrogen lines. Moreover, the
ratio of the peak intensities is greater than the typical interstellar
abundance value of 0.1.

It is worth noting that interstellar absorption lines toward individual
stars in the Carina Nebula have velocity features extending over 550 \kms
(e.g \citeNP{Laurent82} and \citeNP{Walborn84}). There are three main
velocity regimes: a cool foreground gas component at $\sim 10$~\kms; a gas
component at $\sim -30$~\kms; and a very high velocity gas component at
$\sim -100$~\kms. The gas component close to $-30$~\kms\ is thought to be
associated with Car~II. Models for the high-velocity gas component are not
well established. This gas is likely to be from isolated, energetic
wind-driven structures around individual hot stars \cite{Walborn98}. One
explanation for emission components at the two extreme negative velocities
is that they represent H110$\alpha$ emission coming from these wind-driven
structures. Further data are required in order to better identify their
origin.

\section{Conclusion}

We have obtained H110$\alpha$ recombination-line and 4.8-GHz continuum data
on the two bright radio sources in the Carina Nebula --- Car~I and Car~II.

Car~I contains two bright emission arcs, Car~I-E and Car~I-W. The
velocities of the ionized gas associated with each arc are $\sim -19$~\kms\
and $\sim -25$~\kms, respectively. Their ionizing fluxes are consistent
with ionization fronts originating from Tr~14. Car~I-E envelops a dense
fragment at the front of the northern molecular cloud and the new structure
revealed in Car~I-W implies that it envelops a dense molecular fragment in
the middle or back of the northern cloud. Car~I may be expanding and
carving out a cavity within the northern molecular cloud. The electron
temperatures for Car~I range between 3000~K and 8000~K.  Two compact \HII\
regions have been identified with fluxes that correspond to ionization by
single O9.5 and BO-type stars. If this is the case, then this is the first
evidence of ongoing star formation in the northern cloud.  Both sources are
situated in the vicinity of the ionization arcs and could be examples of
triggered star formation.

Unlike Car~I, there is a close resemblance between the main components of
Car~II and the bright features of the optical nebulosity, in particular the
prominent ring-like structure. The kinematics of Car~II are very
complicated. They are consistent with expansion from at least two different
points. There are three main emission components --- Car~II-N is in the
shape of an arc and forms more than half of the ring; and Car~II-E and
Car~II-W form a linear or bar-like feature that extends in the
northeast--southwest direction. Car~II-N is more prominent between
velocities of $-18$ and $-5$~\kms\ and Car~II-E and Car~II-W are more
prominent between velocities of $-42$ and $-34$~\kms. The ionizing fluxes
of these three components are consistent with ionization fronts originating
from the massive stellar members of Tr~16, in particular $\eta$~Car. It is
not clear what is responsible for their striking shapes: Car~I-N may be
enveloping a dense molecular cloud and Car~II-E and Car~II-W may be
bordering a region in the vicinity of $\eta$~Car that has been depleted of
gas.

\section{Acknowledgements}

We thank Neil Killeen, Bob Sault and Miller Goss for assistance with the
data reduction.  We are also grateful to Jim Caswell for helpful
comments. KJB acknowledges the support of an Australian Post-graduate
Award. This work has been supported by a grant from the Australian Research
Council.

\end{document}